%% file: main.tex
\documentclass[manuscript]{acmart}
\usepackage[most]{tcolorbox}
\usepackage{multirow}
\usepackage{forest}

\input{front-matter/packages}

\input{front-matter/acm-specific}

\input{front-matter/blind-unblind-macros}

\begin{document}

\input{front-matter/title}


\input{front-matter/authors}

\input{front-matter/abstract}

%
\newenvironment{changemargin}[2]{%
\begin{list}{}{%
\setlength{\topsep}{6pt}%
\setlength{\leftmargin}{#1}%
\setlength{\rightmargin}{#2}%
\setlength{\parsep}{\parskip}%
}%
\item[]}{\end{list}}

\newcommand{\researchquestion}[1]{%
  \begin{list}{}{%
    \setlength{\topsep}{6pt}%
    \setlength{\leftmargin}{0.25in}%
    \setlength{\rightmargin}{0.25in}%
    \setlength{\parsep}{\parskip}%
  }
  \item[]
    \doublebox{%
      \begin{minipage}{4.75in}
        \em #1
      \end{minipage}%
    }
  \end{list}%
}
\input{front-matter/CCSXML-and-keywords}
%
\maketitle

\input{front-matter/authors-note}

\section{Introduction}
\label{sections-s1-introduction}
\input{sections/s1-introduction}

\label{sections-s2-related-work}
\input{sections/s2-related-work.tex}

\section{Methodology}
\label{sections-s3-method}
\input{sections/s3-method}

\section{Results}
\label{sections-s4-results}
\input{sections/s4-results}

\section{Discussions and Implications}
\label{sections-s5-discussion}
\input{sections/s5-discussion}

\section{Threats to Validity}
\label{sections-s6-threat}
\input{sections/s6-threat}

\section{Conclusion}
\label{sections-s7-conclusion}
\input{sections/s7-conclusion}

\bibliographystyle{ACM-Reference-Format}
\bibliography{bibliography}

\pagebreak
\end{document}

%% file: front-matter/packages.tex
\usepackage{multirow}
\usepackage{listings}
\usepackage{subcaption}
\usepackage[T1]{fontenc}
\usepackage{fancybox,framed}
\usepackage{xcolor}
\usepackage{graphicx}
\usepackage{longtable}
\usepackage{lscape}
\usepackage{todonotes}
\usepackage{multirow}
\usepackage{color}
\usepackage{colortbl}
\definecolor{Gray}{gray}{0.9}
\usepackage[flushleft]{threeparttable}
\usepackage{listings}
\usepackage{enumitem}
\usepackage{tablefootnote}
\usepackage{float}
\usepackage{booktabs}
\usepackage{makecell}

%% file: front-matter/acm-specific.tex
\def\BibTeX{{\rm B\kern-.05em{\sc i\kern-.025em b}\kern-.08emT\kern-.1667em\lower.7ex\hbox{E}\kern-.125emX}}

\acmYear{2025} \acmVolume{1} \acmNumber{1} \acmArticle{1} \acmMonth{1} 
\acmDOI{}

\acmISBN{}

\acmYear{2025} \acmVolume{1} \acmNumber{1} \acmArticle{1} \acmMonth{1} 

\copyrightyear{2025}

%% file: front-matter/blind-unblind-macros.tex
\newcommand{\BLIND}[2]{{\ifdefined\DOUBLEBLIND #2\else #1\fi}}



%% file: front-matter/title.tex
\title{A Systematic Literature Review of the Use of GenAI Assistants for Code Comprehension: Implications for Computing Education Research and Practice}

%% file: front-matter/authors.tex
\ifdefined\DOUBLEBLIND
\input{front-matter/authors-blind}
\else
\input{front-matter/authors-nonblind}
\fi

%% file: front-matter/authors-blind.tex
\author{Author1}
\email{author1@example.org}
\affiliation{%
  \institution{Author1's Department}
  \streetaddress{Author1's Institution}
  \city{City}
  \state{State/Province}
  \country{Country}
  \postcode{postcode}
}

\author{Author2}
\email{author2@example.org}
\affiliation{%
  \institution{Author2's Department}
  \streetaddress{Author2's Institution}
  \city{City}
  \state{State/Province}
  \country{Country}
  \postcode{postcode}
}

\author{Author3}
\email{author3@example.org}
\affiliation{%
  \institution{Author3's Department}
  \streetaddress{Author3's Institution}
  \city{City}
  \state{State/Province}
  \country{Country}
  \postcode{postcode}
}

\author{Author4}
\email{author4@example.org}
\affiliation{%
  \institution{Author4's Department}
  \streetaddress{Author4's Institution}
  \city{City}
  \state{State/Province}
  \country{Country}
  \postcode{postcode}
}

\renewcommand{\shortauthors}{Author1 et al.}

%% file: front-matter/authors-nonblind.tex



\author{Yunhan Qiao}
\email{qiaoy@oregonstate.edu}
\affiliation{%
  \institution{Oregon State University}
  \city{Corvallis}
  \state{Oregon}
  \country{USA}
}

\author{Md Istiak Hossain Shihab}
\email{shihabm@oregonstate.edu}
\affiliation{%
  \institution{Oregon State University}
  \city{Corvallis}
  \state{Oregon}
  \country{USA}
}

\author{Christopher Hundhausen}
\email{chris.hundhausen@oregonstate.edu}
\affiliation{%
  \institution{Oregon State University}
  \city{Corvallis}
  \state{Oregon}
  \country{USA}
}

\renewcommand{\shortauthors}{Qiao et al.}


%% file: front-matter/abstract.tex
\newcommand{\AbstractCategory}[1]{%
  \par\addvspace{.5\baselineskip}
  \noindent\textbf{#1}\quad\ignorespaces
}

%
\begin{abstract}
The ability to comprehend code has long been recognized as an essential skill in software engineering. As programmers lean more heavily on generative artificial intelligence (GenAI) assistants to develop code solutions, it is becoming increasingly important for programmers to comprehend GenAI solutions so that they can verify their appropriateness and properly integrate them into existing code. At the same time, GenAI tools are increasingly being enlisted to provide programmers with tailored explanations of code written both by GenAI and humans. Thus, in computing education, GenAI presents new challenges and opportunities for learners who are trying to comprehend computer programs. To provide computing educators with evidence-based guidance on the use of GenAI to facilitate code comprehension and to identify directions for future research, we present a systematic literature review (SLR) of state-of-the-art approaches and tools that leverage GenAI to enhance code comprehension. Our SLR focuses on 31 studies published between 2022 and 2024. Despite their potential, GenAI assistants often yield inaccurate or unclear explanations, and novice programmers frequently struggle to craft effective prompts, thereby impeding their ability to leverage GenAI to aid code comprehension. Our review classifies GenAI-based approaches and tools, identifies methods used to study them, and summarizes the empirical evaluations of their effectiveness. We consider the implications of our findings for computing education research and practice, and identify directions for future research.
\end{abstract}

%% file: front-matter/CCSXML-and-keywords.tex
\ccsdesc[500]{Code Comprehension}
\ccsdesc[500]{AI Assistants}

\keywords{software engineering education, artificial intelligence}

%% file: front-matter/authors-note.tex

\pagebreak
\textbf{Author's Note:} 
\begin{itemize}
    

    \item The corresponding author may be reached at~\BLIND{\hyperlink{mailto:}{\tt }}{\tt authorone@example.org}
\end{itemize}

%% file: sections/s1-introduction.tex
Code comprehension is the process by which programmers extract useful information from computer code to construct a mental model representing their understanding of the code \cite{graesser2013prose}. 
It is crucial for software development. At the level of code blocks, functions, and modules, it aids in understanding program logic and functionality. At the level of software systems, comprehension is essential for software maintenance and evolution, accounting for approximately 52\% to 70\% of the time required for those activities \cite{al2017source, koenemann1991expert, fjeldstad1983application, xia2017measuring}.

The emergence of Large Language Models (LLMs) in artificial intelligence has led to a proliferation in so-called Generative AI (GenAI) assistants (e.g. Github Copilot \cite{github_copilot}) in both the software industry and computer science education \cite{geng2024large,smith2024prompting,sarsa2022automatic,geng2022fine, jiang2406survey, nam2024using}. These GenAI assistants have shifted the focus of coding tasks from writing code to comprehending, refining, and integrating LLM-generated code \cite{denny2024explaining,nam2024using}. While GenAI assistants can produce innaccurate or difficult-to-understand explanations, programmers often request that GenAI assistants explain specific parts of both existing code, and code generated by GenAI assistants. 
Moreover, instructors are increasingly using GenAI assistants to automatically generate diverse code explanations to support student learning \cite{macneil2022generating,sarsa2022automatic}. Additionally, GenAI assistants are increasingly being enlisted to assess students' comprehension skills \cite{macneil2023experiences, nam2024using, choudhuri2024far}. Thus, GenAI assistants are playing an increasingly important role in computer programming and computing education both by transforming coding tasks into comprehension activities, and by producing code explanations.

Although numerous studies have explored ways to improve code comprehension with GenAI assistants, there are no comprehensive reviews of current state-of-the-art (SOTA) approaches and tools for enhancing code comprehension through GenAI assistants. This gap may hinder research and practical application, as the absence of a comprehensive review of the effectiveness of existing GenAI tools can make it challenging for both for researchers to identify current trends, methods, and unresolved challenges, and for students and teachers to adopt the most effective strategies for computing education.

To address this gap, we present a systematic literature review (SLR) on SOTA approaches, tools, and studies related to the use of GenAI assistants for code comprehension. Following the methodology of prior SLRs (e.g., \cite{feng2024guiding}), we analyzed 31 papers published between 2022 and 2024, guided by the following research questions:
\begin{itemize}
    \item \textbf{RQ1: }How can GenAI assistants facilitate code comprehension?
    \item \textbf{RQ2: }What methods have been used to study the use of GenAI in facilitating code comprehension?
    \item \textbf{RQ3: }How effective are GenAI assistants in facilitating code comprehension?
   
\end{itemize}
In addressing these questions, we make two contributions to the field: (1) we summarize the SOTA approaches and tools that employ GenAI assistants to enhance code comprehension, including the methods used to evaluate them and their effectiveness; and (2) we identify the implications of our findings for computing education research and practice.

%% file: sections/s2-related-work.tex
\section{Related Work}
\subsection{Code Comprehension}
Code comprehension is the process by which programmers construct mental models from code snippets or entire code bases \cite{heinonen2023synthesizing}. It typically involves using either top-down \cite{brooks1977towards} or bottom-up \cite{letovsky1987cognitive} approaches. Top-down approaches start with the high-level abstractions, structure and organization of a code base and then move to its lower-level components, whereas bottom-up approaches start with low-level components and move to higher-level organization and abstractions. 

In practice, programmers often switch between these approaches based on their background and experience. Novices, who generally lack extensive background knowledge, tend to rely on a bottom-up approach to understand program semantics because they find it challenging to grasp the abstract features of the program \cite{cornelissen2010controlled}. In contrast, experienced developers are more likely to adopt a top-down approach, as they can efficiently recognize beacons, cues, and familiar programming patterns \cite{teasley1994effects}. Additionally, programmers apply various practices to extract information from code, such as reading code line by line, mentally analyzing it, or running and debugging it \cite{heinonen2023synthesizing}. During the comprehension process, programmers often generate hypotheses about the functionality, features, and structure of the program, confirming or discarding those hypotheses as their understanding deepens \cite{von1995program}.

In software development, code comprehension plays a vital role in enabling software developers to better maintain and evolve their code \cite{tripathy2014software}. Understanding a software system's code and structure is the first step in performing the five central software maintenance and evolution tasks: \textit{adaptive}, \textit{perfective}, \textit{corrective}, \textit{rescue}, and \textit{code leverage} \cite{tripathy2014software, levy2019understanding}. Code comprehension is also time consuming: According to several empirical studies, comprehension accounts for approximately 52\% to 70\% of maintenance and evolution time \cite{al2017source, koenemann1991expert, fjeldstad1983application, xia2017measuring}.

\subsection{Evaluating Code Comprehension} 
Researchers have long been interested in studying how programmers understand different aspects of code. To that end, at least four different evaluation methods have been employed. The most popular method is asking participants to provide information about the program (\cite{oliveira2020evaluating, wyrich202340}. For instance, Baron et al. \cite{baron2024recursion} conducted an experiment where participants were asked to provide short descriptions of the functionality of recursion code snippets. Another way to assess code comprehension is by asking participants to provide personal opinions based on their understanding of a program \cite{oliveira2020evaluating}. For example, Cates et al. \cite{cates2021does} asked participants to suggest better names for anonymous functions based on their understanding of the functions. Third, it is common to ask subjects to provide subjective ratings of their confidence in their comprehension \cite{santos2022jask, stapleton2020human}. A fourth method is to have programmers perform code edits based on their understanding of code \cite{oliveira2020evaluating}. For instance, Hofmeister et al. \cite{hofmeister2017shorter} asked participants to find and correct defects in provided code snippets. 

In addition, some studies have evaluated the performance of GenAI models in code comprehension tasks (i.e., code understanding ability) across different code understanding benchmarks \cite{zeng2022extensive, li2023evaluating, zhang2024detecting, sahu2024codequeries, li2024mutation, lehtinen2024let}. Likewise, some studies have evaluated the readability of GenAI-generated code \cite{nguyen2022empirical, haindl2024does, al2022readable}. In contrast to these efforts, in this paper we focus on papers that \textit{evaluate the effectiveness} GenAI in helping people comprehend code.



\subsection{Code Comprehension with GenAI Assistants} 
With the boom in LLMs, it is crucial for novices and professional developers to comprehend the code generated by GenAI assistants. At the same time, GenAI assistants can be leveraged to help developers comprehend code written by humans \cite{denny2024explaining,nam2024using}. Research indicates that explanations from GenAI assistants generally cover most of the code but sometimes include inaccuracies \cite{sarsa2022automatic}. Nonetheless, students show a preference for explanations generated by GenAI assistants over their own. This preference arises because students often lack the skills to create better explanations, especially for introductory computer science (CS) tasks, where understandability and readability are crucial. \cite{leinonen2023comparing}. 

Previous studies have been interested in understanding the impact of GenAI assistance on programming efficiency and effectiveness \cite{peng2023impact, pandey2024transforming, ziegler2024measuring, kalliamvakou2022research, dohmke2023sea, smit2024impact}. Relatively little work has focused on comprehension tasks.
Some recent studies focused on users' comprehension of code generated by GenAI assistants during basic programming tasks, such as introductory computer science assignments or small-scale programs in novel domains. The results have been mixed. In web development classes, students favored line-by-line explanations when working with Copilot \cite{macneil2023experiences}. In more complex software engineering (SE) tasks, using GenAI assistants significantly reduced task completion time, but did not improve students' code understanding abilities \cite{nam2024using}. Moreover, students experienced frustration due to their incorrect comprehension of GenAI assistants' responses \cite{choudhuri2024far}. To enhance AI assistants' code explanations, Nam et al. \cite{nam2024using} developed an IDE plugin that aids users in comprehending unfamiliar APIs through the support of AI assistants. Similarly, Yan et al. \cite{yan2024ivie} created a tool that generates anchored explanations with AI assistants to help users understand AI-generated code.

%% file: sections/s3-method.tex
\begin{figure}[h!]
    \centering
    \includegraphics[width=\textwidth]{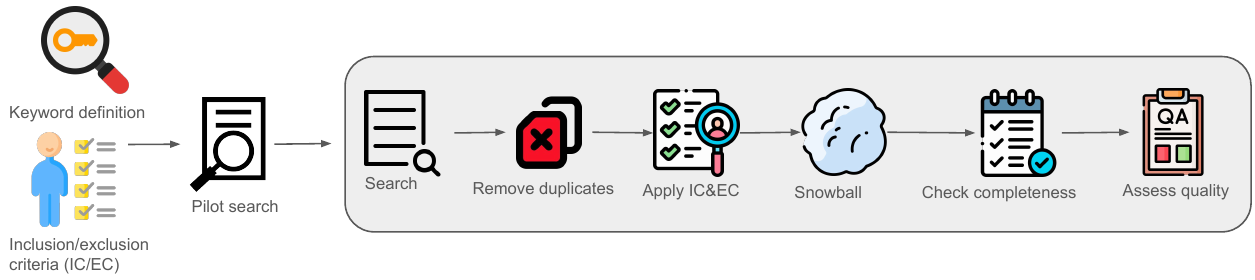}
    \caption{SLR Methodology}
    \label{fig:SLR}
\end{figure}

We conducted a systematic literature review (SLR) following an established methodology used in previous studies SLRs in the computing discipline \cite{feng2024guiding,kitchenham2013systematic}. Figure \ref{fig:SLR} presents the methodology, which includes defining the search string, searching for papers, removing duplicates, filtering with include/exclude criteria, snowballing, checking for completeness, and assessing quality.

Following the guidance of Feng et al. \cite{feng2024guiding}, we developed a data collection strategy that searched three prominent digital libraries: ACM Digital Library, IEEE Xplore, and ScienceDirect. The selected digital libraries are frequently used in other SLR studies in computing \cite{feng2024guiding, stol2016grounded, heinonen2023synthesizing, yang2024decoding}. We decided not to filter papers based on the publication year, as studies of GenAI have emerged only in the past few years. 

\subsection{Search String Definition}
Following the study by Yang et al. \cite{yang2024decoding}, we identified search keywords relative to three dimensions: comprehension, code, and AI. For the comprehension and code dimensions, we created a word cloud based on two foundational papers on code comprehension \cite{von1995program, letovsky1987cognitive} and one paper on computer science education \cite{robins2003learning}. For the AI dimension, we added additional keywords we deemed relevant. This resulted in the following keywords:
\begin{itemize}
    \item \textbf{Comprehension:} comprehen*, understand*, expla*
    \item \textbf{Code:} program*, cod*, software
    \item \textbf{AI}: gpt, AI, LLM, Large Language Model, copilot
\end{itemize}

We designed the search strings to ensure comprehensive coverage of relevant studies by requiring the terms to appear at least once in the paper's title, abstract, or author keywords. In the ACM Digital Library and IEEE Explore, the search strings were applied to the title, abstract, and author keywords fields of the search. In ScienceDirect, we searched using the "Title, Abstract, or Author-Specified Keywords" search box. Since ScienceDirect does not support wildcard characters, we used the exact terms "comprehend," "explain," "understand," "code," and "program" for our searches. The complete search strings used for ACM and IEEE are shown in Table \ref{tab: search string for acm}, while the search strings for ScienceDirect are shown in Table \ref{tab: search string for SD}.

\begin{table}[ht]
\centering
\caption{Search Strings for ACM and IEEE}
\small
\begin{tabular}{c p{10.5cm}}
\hline
\textbf{Search String} & \textbf{Query} \\
\hline
S1 & abstract: (comprehen* OR understand* OR expla*) AND (AI OR gpt OR Copilot OR LLM OR "Large Language Model") AND (cod* OR program* OR software) \\
S2 & title: (comprehen* OR understand* OR expla*) AND (AI OR gpt OR Copilot OR LLM OR "Large Language Model") AND (cod* OR program* OR software) \\
S3 & author keywords: (comprehen* OR understand* OR expla*) AND (AI OR gpt OR Copilot OR LLM OR "Large Language Model") AND (cod* OR program* OR software) \\
\hline
\end{tabular}
\label{tab: search string for acm}
\end{table}

\begin{table}[ht]
\centering
\caption{Search Strings for ScienceDirect}
\small
\begin{tabular}{c p{10.5cm}}
\hline
\textbf{Search String} & \textbf{Query} \\
\hline
S1 & (comprehend OR understand OR explain) AND AI AND (code OR program* OR software) \\
S2 & (comprehend OR understand OR explain) AND gpt AND (code OR program OR software) \\
S3 & (comprehend OR understand OR explain) AND Copilot AND (cod* OR program* OR software) \\
S4 & (comprehend OR understand OR explain) AND LLM AND (cod* OR program* OR software) \\
S5 & (comprehend OR understand OR explain) AND "Large Language Model" AND (cod* OR program* OR software) \\
\hline
\end{tabular}
\label{tab: search string for SD}
\end{table}

\subsubsection{Pilot Study}
We conducted a pilot search on three databases to validate our search strategy and ensure relevance to the research questions. We selected three papers \cite{nam2024using,yan2024ivie,geng2024large} involving GenAI assistants published in top software engineering and HCI conferences. We successfully retrieved all three papers using the defined search strategy, so no further changes were made to the search string.

\subsection{Eligibility Criteria}
Following previous software engineering SLR studies \cite{feng2024guiding, stol2016grounded, heinonen2023synthesizing}, we designed inclusion criteria (IC) and exclusion criteria (EC) to narrow down the preliminary search results to those relevant to our research questions, Table \ref{tab: IC&EC} presents those criteria.
\begin{table}[ht]
\centering
\caption{Inclusion and Exclusion Criteria}
\small
\begin{tabular}{c p{10.5cm}}
\hline
\textbf{Criterion} & \textbf{Description} \\
\hline
IC 1 & The paper studies a human's ability to comprehend GenAI-generated code and/or explanations of code. \\
IC 2 & The paper studies the effectiveness of GenAI tools in helping humans comprehend code written by others. \\
EC 1 & The paper is a doctoral/master's thesis, book, research proposal, poster, blog, keynote, invited talk, abstract, demonstration, or technical paper. \\
EC 2 & The paper is not accessible. \\
EC 3 & The paper is not peer reviewed. \\
EC 4 & The paper is not written in English. \\
EC 5 & The paper focuses on GenAI training and fine-tuning and the technical details of GenAI. \\
\hline
\end{tabular}
\label{tab: IC&EC}
\end{table}

\begin{table}
\centering
\caption{Initial hits in three digital libraries}
\small
\begin{tabular}{ p{1.1cm} p{1cm} p{1cm} p{1.4cm} p{1.2cm} p{1cm}}
\hline
 \textbf{Search Strings} & \textbf{ACM} & \textbf{IEEE Xplore} & \textbf{Science Direct}\\ 
\hline
 S1 &  1566 & 1683 & 557 \\  
 S2 & 23 & 17 & 25  \\
 S3 & 39  & 66 & 2  \\
 S4 & N/A & N/A & 25 \\
 S5 & N/A & N/A & 167 \\
 \hline
 \textbf{Total} & 1628 & 1766 & 776 \\
\hline
\end{tabular}
\label{tab:paper_counts}
\end{table}

\subsection{SLR Procedure}
\subsubsection{Searching Papers and Removing Duplicates}
Table \ref{tab:paper_counts} presents the initial number of papers retrieved by our searches in the three digital libraries. We searched on December 6th, 2024. After identifying and removing duplicates across the different libraries, we removed 240, resulting in 3930 unique papers.

\subsubsection{Applying IC and EC}
Before applying the IC and EC, the paper's first and second authors sampled 100 papers and discussed their eligibility based on the IC and EC. Then, we divided the papers equally between the first and second authors. We read each paper's title, abstract, and introduction to understand its purpose and contributions. We discussed whether to keep or remove a paper if we were uncertain. If we could not reach an agreement, we kept the paper. After applying the IC and EC, we retained 58 eligible papers.

\subsubsection{Snowballing}
 Following the guidance of prior studies \cite{feng2024guiding,stol2016grounded,heinonen2023synthesizing}, we conducted a single round of backward and author-based snowballing to attain better coverage. During backward snowballing, we evaluated the references of each eligible paper based on its abstract and introduction. For author-based snowballing, we reviewed the work of three prominent researchers in human-computer interaction, empirical software engineering, and computer science education: Brad A. Myers, Margaret-Anne Storey, and Paul Denny. We examined their Google Scholar profiles to identify relevant papers they had authored, continuing the search until no additional relevant papers were found. As a result, backward snowballing retrieved 13 additional relevant papers, while author-based snowballing yielded three relevant papers. After snowballing, we had 74 eligible papers.

\subsubsection{Checking Completeness}
To ensure that our paper search was comprehensive, we conducted a completeness check on the four most cited papers in our list after snowballing \cite{sarsa2022automatic, kazemitabaar2023studying, nguyen2022empirical, ross2023programmer}. After reviewing all the references in each paper, we retrieved one additional eligible paper. Thus, our completeness rate was 74 out of 75, or 98.7\%.

\subsubsection{Quality Assessment}
To ensure that all selected papers were relevant to the research questions, we evaluated the full text of each paper based on a list of defined quality assessment criteria (AC) shown in Table \ref{tab:QAC}. Observe that these assessment criteria exclude papers that empirically study how students' or developers' comprehend GenAI-generated code better (e.g., \cite{zeng2022extensive, li2023evaluating, zhang2024detecting, sahu2024codequeries, li2024mutation, lehtinen2024let,nguyen2022empirical, haindl2024does, al2022readable}). Instead, the criteria focus exclusively on approaches that use the GenAI to enhance code comprehension. After we applied the AC, we obtained 31 papers.

\begin{table}[ht]
\centering
\caption{Assessment criteria (AC) definition}
\small
\begin{tabular}{c p{2.5cm} p{8cm}}
\hline
\textbf{ID} & \textbf{Criterion} & \textbf{Criterion definition} \\ \hline
AC1 & Pedagogy & The paper explicitly proposes a novel pedagogy using AI assistants to support students' code comprehension. \\ 
AC2 & Tools/approaches & The paper presents a tool/approach that enhances students' or developers' comprehension of code. \\\hline
\end{tabular}
\label{tab:QAC}
\end{table}

\subsection{Data Extraction Coding}
Following the guidance of previous studies \cite{feng2024guiding,yang2024decoding}, we employed an open-coding protocol to systematically extract text segments from these papers that address our research questions. In this process, each code was compared to previously identified categories to determine whether it belonged to an existing category or represented a new one. We did this until all the codes were saturated.

To verify the reliability of our categorizations, the first two authors independetly applied our coding schemes to a 20\% sample of the papers. We then calculated the inter-rater reliability score using Krippendorff's Alpha. Table \ref{tab:RQ2_summary} presents inter-rater reliability scores for each set of categories. As Table \ref{tab:RQ2_summary} suggests, the scores ranged from 0.70 to 0.81, indicating acceptable to good agreement. Having established sufficient inter-rater reliability, we completed the coding by equally dividing the remaining 80\% of the papers between the first and second authors.

\begin{table}[h!]
\centering
\footnotesize
\caption{Krippendorff's Alpha Scores for Inter-Rater Reliability (IRR)}
\label{tab:RQ2_summary}
\begin{tabular}{lll}
\hline
\textbf{Measures}                   & \textbf{Categories}                              & \textbf{IRR}            \\ \hline
\multirow{4}{*}{Dependent Measures (DM)} & Percept (Per)      & \multirow{5}{*}{0.81} \\
                                    & Process Metrics (PrM)                              &                         \\
                                    & Performance Metrics (PeM)                           &                         \\
                                    & NLP Metrics (NM)                                     &                         \\ \hline
\multirow{5}{*}{Data Collection Methods (DCM)} & Survey/Questionnaire (SQ)                            & \multirow{5}{*}{0.79}                     \\
                                    & Field Study (FS)                             &                      \\
                                    & Lab Study (LS)                              &                     \\
                                    & Interview (I)                                       &                      \\
                                    & Interaction Logging (IL)                             &                      \\ \hline
\multirow{4}{*}{Data Analysis Methods (DAM)} & Qualitative Analysis (QLA)                            & \multirow{4}{*}{0.71} \\
                                  & Quantitative Analysis (QTA) & \\
                                  & Mixed Method (MM) & \\ \hline
\multirow{4}{*}{Participants (P)}        & Students (S)                                        & \multirow{4}{*}{0.70}                     \\
                                    & Researchers (R)                                     &                     \\
                                    & Educators (E)                                       &                     \\
                                    & Professionals (P)                                   &                     \\ \hline
\end{tabular}
\end{table}

%% file: sections/s4-results.tex
\begin{figure}[h!]
    \centering
    \caption{Number of papers published by year}    \includegraphics[width=0.6\textwidth]{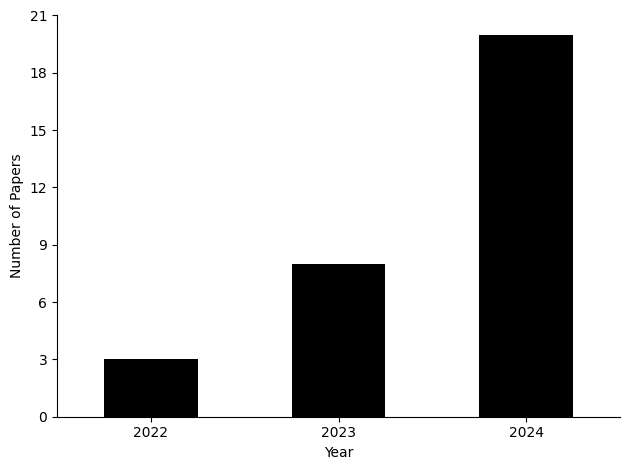}
    \label{fig:papers_by_year}
\end{figure}
\begin{figure}[h!]
    \centering
    \caption{Number of papers published by conference/journal}
    \includegraphics[width=0.7\textwidth]{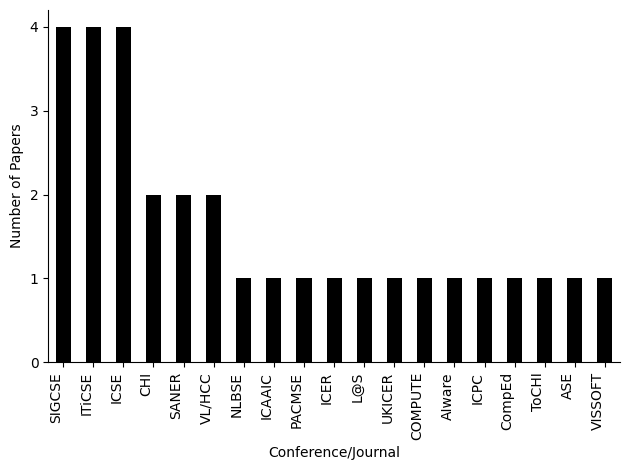}
    \label{fig:papers_by_con}
\end{figure}
\begin{figure}[h!]
    \centering
    \includegraphics[width=0.7\textwidth]{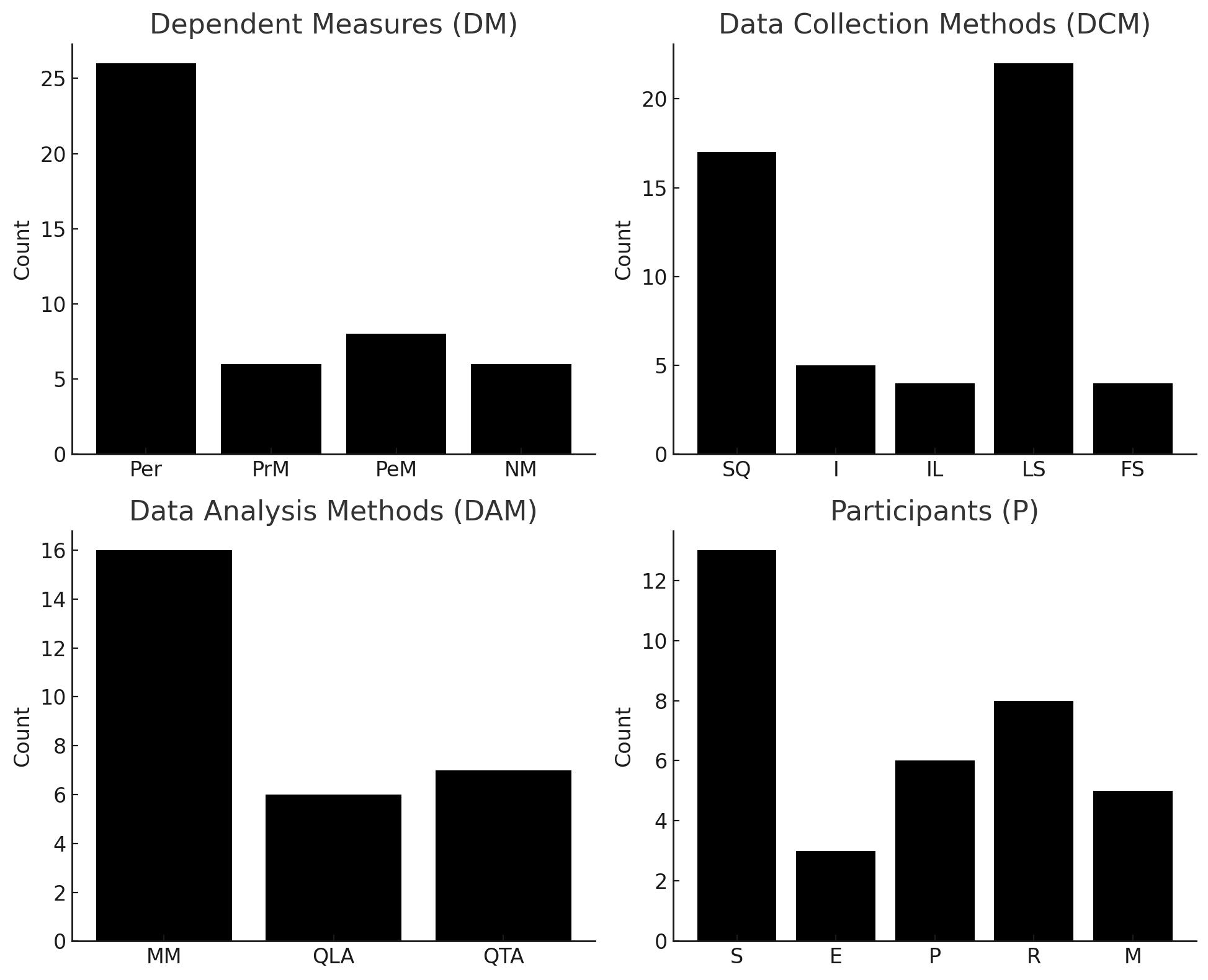}
    \caption{Counts of Category Items}
    \label{fig:countsofcategory}
\end{figure}

Figures \ref{fig:papers_by_year} and \ref{fig:papers_by_con} present the counts of articles by publication year and conference/journal. All papers were published between 2022 and 2024, coinciding with the recent emergence of GenAI assistants. Moreover, the number of papers published has been increasing, indicating that leveraging GenAI to address code comprehension issues is attracting increasing research attention. Overall, the mean number of citations for the 31 papers is 52.65 (\textit{SD} = 89.74). Table \ref{tab:master} presents the categories for the 31 papers, while Figure \ref{fig:countsofcategory} presents the number of papers classified into each category.

\begin{landscape}
\setlength{\tabcolsep}{3pt}
\renewcommand{\arraystretch}{0.9}

\begin{table}[ht!]
  \centering
  \tiny
  \caption{%
    Summary of all 31 papers in terms of their Approach, Dependent Measures (DM), 
    Data Collection Methods (DCM), Data Analysis Methods (DAM), and Participants (P).  
    The abbreviations are defined in Table~\ref{tab:abbr-legend}.%
  }
  \label{tab:master}

  \begin{tabular}{|l| *{10}{c|} *{5}{c|} *{5}{c|} *{3}{c|} *{5}{c|}}
    \hline

    \multirow{2}{*}{\textbf{Paper}}
      & \multicolumn{5}{c|}{\textbf{Approach}}
      & \multicolumn{4}{c|}{\textbf{Dependent Measures}}
      & \multicolumn{5}{c|}{\textbf{Data Collection Methods}}
      & \multicolumn{3}{c|}{\textbf{Data Analysis Methods}}
      & \multicolumn{5}{c|}{\textbf{Participants}} \\
    \cline{2-23}
& \textbf{ExpSoft}
& \textbf{EnhPed}
& \textbf{EnhDoc}
& \textbf{EnRead}
& \textbf{VisSoft}
    & \textbf{Per}
    & \textbf{PrM}
    & \textbf{PeM}
    & \textbf{NM}
    & \textbf{SQ}
    & \textbf{I}
    & \textbf{IL}
    & \textbf{LS}
    & \textbf{FS}
    & \textbf{MM}
    & \textbf{QLA}
    & \textbf{QTA}
    & \textbf{S}
    & \textbf{E}
    & \textbf{P}
    & \textbf{R}
    & \textbf{M} \\
    \hline

    \cite{khan2023combining}
    &   &   &  {$\checkmark$} &   &   
    & {$\checkmark$} &   &   &    
    & {$\checkmark$} &   &   & {$\checkmark$} &   
    &   &   & {$\checkmark$} 
    &   &   &   &   & {$\checkmark$} \\ \hline 

    \cite{nam2023improving}
    &   &   &  {$\checkmark$} &   &   
    & {$\checkmark$} & {$\checkmark$} & {$\checkmark$} &   
    &   & {$\checkmark$} &   & {$\checkmark$} &   
    & {$\checkmark$} &   &   
    &   &   & {$\checkmark$} &   &   \\ \hline 

    \cite{wang2022documentation}
    &   &   &  {$\checkmark$} &   
    & {$\checkmark$} & {$\checkmark$} & {$\checkmark$} & {$\checkmark$} &  
    &   &   &   & {$\checkmark$} &  
    & {$\checkmark$} &   &  
    &   &   &   & {$\checkmark$} &   \\ \hline

    \cite{kazemitabaar2024codeaid}
    & {$\checkmark$} &   &   &   &   
    & {$\checkmark$} & {$\checkmark$} &   &   
    & {$\checkmark$} & {$\checkmark$} & {$\checkmark$} &   &   
    & {$\checkmark$} &   &   
    & {$\checkmark$} & {$\checkmark$} &   &   &   \\ \hline 

    \cite{yan2024ivie}
    & {$\checkmark$} &   &   &   &   
    & {$\checkmark$} &  & {$\checkmark$} & 
    & {$\checkmark$} & {$\checkmark$} & {$\checkmark$} & {$\checkmark$} & {$\checkmark$}
    & {$\checkmark$} &   & 
    & {$\checkmark$} &   &   &   &   \\ \hline 

    \cite{nam2024using}
    & {$\checkmark$} &   &   &   &  
    &   & {$\checkmark$} & {$\checkmark$} &  
    & {$\checkmark$} &   &   & {$\checkmark$} &  
    & {$\checkmark$} &   &  
    & {$\checkmark$} &   & {$\checkmark$} &   &   \\ \hline 

    \cite{tang2024towards}
    & {$\checkmark$} &   &   &   &  
    &   &   &   &   
    &   &   &   &   &   
    &   &   &   
    &   &   &   &   &   \\ \hline 

    \cite{balfroid2024towards}
    & {$\checkmark$} &   &   &   &   
    & {$\checkmark$} &   &   &  
    &   &   &   & {$\checkmark$} &   
    & {$\checkmark$} &   & 
    &   &   &   & {$\checkmark$} &   \\ \hline

    \cite{nazari2024generating}
    &   &   &   & {$\checkmark$} &   
    & {$\checkmark$} &   &   &   
    & {$\checkmark$} &   &   & {$\checkmark$} &  
    & {$\checkmark$} &   &  
    & {$\checkmark$} &   &   &   &   \\ \hline 

    \cite{bernstein2024like}
    &   & {$\checkmark$}  &   &   &  
    & {$\checkmark$} &   &   & 
    & {$\checkmark$} &   &   &   & {$\checkmark$} 
    & {$\checkmark$} &   &   
    & {$\checkmark$} &   &   &   &   \\ \hline

    \cite{denny2024explaining}
    &   & {$\checkmark$} &   &  &  
    & {$\checkmark$} &   & {$\checkmark$} & 
    &   &   &   &   & {$\checkmark$} 
    & {$\checkmark$} &   &   
    & {$\checkmark$} &   &   &   &   \\ \hline 

    \cite{smith2024prompting}
    &   & {$\checkmark$} &   &   & 
    & {$\checkmark$} &   & {$\checkmark$} & 
    & {$\checkmark$}&   &   &   & {$\checkmark$} 
    & {$\checkmark$} &   &  
    & {$\checkmark$} &   &   &   &   \\ \hline 

    \cite{ahmed2024automatic}
    & {$\checkmark$} &   &   &   &   
    &   &   &   & {$\checkmark$}  
    &   &   &   & {$\checkmark$} &  
    &   &   & {$\checkmark$}  
    &   &   &   &   & {$\checkmark$} \\ \hline 

    \cite{geng2024large}
    & {$\checkmark$} &   &  &   &   
    & {$\checkmark$} &   &   & {$\checkmark$} 
    & {$\checkmark$} &   &   & {$\checkmark$} &   
    &   &   & {$\checkmark$} 
    & {$\checkmark$} &   &   & {$\checkmark$} &   \\ \hline 

    \cite{liu2024interpretable}
    & {$\checkmark$} &   &   &   &   
    & {$\checkmark$} &   &   & {$\checkmark$}  
    & {$\checkmark$} &   &   & {$\checkmark$} &   
    & {$\checkmark$} &   &  
    &   &   & {$\checkmark$} &   & {$\checkmark$} \\ \hline 

    \cite{leinonen2023using}
    & {$\checkmark$} &   &  &   &  
    & {$\checkmark$} &   &   &   
    &   &   &   & {$\checkmark$} &   
    &   & {$\checkmark$} &  
    &   &   &   & {$\checkmark$} &   \\ \hline 

    \cite{amburle2024ai}
    & {$\checkmark$} &   &   &   &   
    & {$\checkmark$} &   & & {$\checkmark$} 
    & {$\checkmark$} &   &   & {$\checkmark$} &   
    &   &   & {$\checkmark$} 
    &   &   & {$\checkmark$} &   &   \\ \hline 

    \cite{widyasari2024demystifying}
    & {$\checkmark$} &   &   &   &   
    & {$\checkmark$} &   &   & {$\checkmark$}  
    & {$\checkmark$} &   &   & {$\checkmark$} &   
    & {$\checkmark$} &   &   
    &   &   & {$\checkmark$} &   &   \\ \hline

    \cite{taylor2024dcc}
    & {$\checkmark$} &   &  &   &   
    & {$\checkmark$} & {$\checkmark$} &   &  
    &   &   & {$\checkmark$} &   & 
    &   & {$\checkmark$} &  
    &   &   &   & {$\checkmark$} &   \\ \hline 

    \cite{santos2024not}
    & {$\checkmark$} &   &   &   &  
    & {$\checkmark$} &   & {$\checkmark$} &  
    &   &   &   & {$\checkmark$} &   
    & {$\checkmark$} &   &   
    & {$\checkmark$} &   &   &   &   \\ \hline 

    \cite{balse2023evaluating}
    & {$\checkmark$} &   &   &   &  
    & {$\checkmark$} &   &   &  
    & {$\checkmark$} &   &   & {$\checkmark$} &   
    &   &   & {$\checkmark$} 
    & {$\checkmark$} &   &   &   &   \\ \hline 

    \cite{santos2023always}
    & {$\checkmark$} &   &   &   &   
    & {$\checkmark$} &   &   & 
    & {$\checkmark$} &   &   & {$\checkmark$} &   
    &   & {$\checkmark$} &   
    &   &   &   & {$\checkmark$} &   \\ \hline

    \cite{macneil2023experiences}
    & {$\checkmark$} &   &   &   &   
    & {$\checkmark$} & {$\checkmark$} & {$\checkmark$} & 
    & {$\checkmark$} &   & {$\checkmark$} &   & 
    & {$\checkmark$} &   &   
    & {$\checkmark$} &   &   &   &   \\ \hline 

    \cite{kang2024quantitative}
    & {$\checkmark$} &   &   &   &  
    & {$\checkmark$} &   &   &  
    &   & {$\checkmark$} &   & {$\checkmark$} &   
    & {$\checkmark$} &   &  
    &   &   & {$\checkmark$} &   &   \\ \hline 

    \cite{cucuiat2024feedback}
    & {$\checkmark$} &   &   &   &  
    & {$\checkmark$} &   &   &
    &   & {$\checkmark$} &   &   &   
    &   & {$\checkmark$} &   
    &   & {$\checkmark$} &   & {$\checkmark$} &   \\ \hline 

    \cite{sarsa2022automatic}
    & {$\checkmark$} &   &   &   &   
    & {$\checkmark$} &   &   & 
    &   &   &   & {$\checkmark$} &   
    &   & {$\checkmark$} &   
    &   &   &   &   & {$\checkmark$} \\ \hline 

    \cite{leinonen2023comparing}
    & {$\checkmark$} &   &  &   &   
    & {$\checkmark$} &   &   & 
    & {$\checkmark$} &   &   & {$\checkmark$} &   
    & {$\checkmark$} &   &   
    & {$\checkmark$} &   &   &   &   \\ \hline 

    \cite{nguyen2024using}
    & {$\checkmark$} &   &   &   &  
    & {$\checkmark$} &   &   &  
    & {$\checkmark$}&   &   & {$\checkmark$} &  
    &   & {$\checkmark$} &   
    & {$\checkmark$} & {$\checkmark$} &   &   &   \\ \hline 

    \cite{khan2022automatic}
    &   &   &{$\checkmark$} &   &   
    &   &   &   & {$\checkmark$} 
    &   &   &   & {$\checkmark$} &   
    &   &   & {$\checkmark$} 
    &   &   &   &   & {$\checkmark$} \\ \hline 

    \cite{dvivedi2024comparative}
    &   &   & {$\checkmark$} &   &   
    & {$\checkmark$} &   &   & 
    & {$\checkmark$} &   &   & {$\checkmark$} &  
    &   &   & {$\checkmark$} 
    &   &   &   & {$\checkmark$} &   \\ \hline 

    \cite{heidrich2023visualizing}
    &   &   &   &   & {$\checkmark$} 
    &   &   &   &   
    &   &   &   &   &  
    &   &   &   
    &   &   &   &   &   \\ \hline 

    \hline
    \textbf{Count}
      & 21   
      & 3   
      & 5   
      & 1   
      & 1   
      & 26  
      & 6   
      & 8   
      & 6   
      & 17  
      & 5   
      & 4   
      & 22  
      & 4   
      & 16  
      & 6   
      & 7  
      & 13  
      & 3   
      & 6   
      & 8   
      & 5 \\ 
    \hline

  \end{tabular}
\end{table}
\end{landscape}

\begin{table}[ht]
  \centering
  \tiny
  \caption{Explanation of Abbreviations used in Table~\ref{tab:master}}
  \label{tab:abbr-legend}
  \rowcolors{2}{gray!15}{white}%
  \resizebox{\textwidth}{!}{%
    \begin{minipage}{\textwidth}
      \centering
      \begin{tabular}{p{1.5cm} p{2cm} p{3cm} p{6.4cm}}
        \toprule
        \multicolumn{4}{l}{\textbf{Approach to Assessing or Fostering Code Comprehension with GenAI}} \\
        \midrule
        \textbf{Abbreviation} & \textbf{Description}         & \textbf{Subcategory}               & \textbf{Description}    \\
        ExpSoft                & Explain Software             & Explain Code                       & Explain code snippet by using GenAI to generate explanation                                      \\
                               &                              & Explain Stack Traces               & Explain stack traces by using GenAI to generate explanation                                       \\
                               &                              & Explain Error Message              & Using GenAI to enhance error messages                                                             \\
                               &                              & Explain Logs                       & Using GenAI to interpret the software logs                                                        \\
        EnhDoc                 & Enhance Documentation        & Enrich Documentation with ML       & Using ML to extract comparable APIs from StackOverflow to supplement code documentation            \\
                               &                              & Generate Executable Code Examples  & Using GenAI to generate executable code examples to supplement code documentation                  \\
                               &                              & Auto-generate Documentation        & Using GenAI models to automatically generate documentation                                        \\
        EnhPed                 & Enhance Pedagogy             & Explain Code with Purpose          & Explaining the code with purpose, then feeding it to GenAI to generate similar code               \\
                               &                              & Generate Analogy                   & Using GenAI to generate analogies to aid code understanding                                       \\
        VisSoft                & Visualize Software           & —                                  & Develop and evaluate GenAI-generated software visualizations to aid code comprehension             \\
      \end{tabular}

      \vspace{1em}

      \begin{tabular}{p{2.5cm} p{11cm}}
        \toprule
        \multicolumn{2}{l}{\textbf{Dependent Measures (DM)}}               \\
        \midrule
        \textbf{Abbreviation} & \textbf{Description}                                        \\
        Per                   & Percept--Users’ impressions of clarity, correctness, relevance, usefulness, comprehensiveness of AI outputs  \\
        PrM                   & Process Metrics--Participants’ behavior when interacting with GenAI tools                                      \\
        PeM                   & Performance Metrics--Participants’ performance in lab studies                                                    \\
        NM                    & NLP Metrics--Using NLP metrics to assess GenAI-generated explanation quality                   \\
        \midrule
        \multicolumn{2}{l}{\textbf{Data Collection Methods (DCM)}}                            \\
        \midrule
        \textbf{Abbreviation} & \textbf{Description}                                        \\
        SQ & Survey / Questionnaire                                                            \\
        IL & Interaction Logging (within the development environment)                           \\
        I  & Interview                                                                         \\
        LS & Lab Study                                                                         \\
        FS & Field Study                                                                       \\
        \midrule
        \multicolumn{2}{l}{\textbf{Data Analysis Methods (DAM)}}                             \\
        \midrule
        \textbf{Abbreviation} & \textbf{Description}                                        \\
        MM  & Mixed Methods (quantitative + qualitative)                                      \\
        QLA & Qualitative Analysis (e.g., thematic coding)                                     \\
        QTA & Quantitative Analysis (e.g., statistical testing)                                \\
        \midrule
        \multicolumn{2}{l}{\textbf{Participants (P)}}                                      \\
        \midrule
        \textbf{Abbreviation} & \textbf{Description}                                        \\
        S & Students                                                                          \\
        E & Educators                                                                         \\
        P & Professionals                                                                     \\
        R & Researchers                                                                       \\
        M & GenAI Models                                                                      \\
        \bottomrule
      \end{tabular}
    \end{minipage}%
  }
\end{table}

In the remainder of this section, we organize our results around the three research questions we posed for this study.

\subsection{RQ1: \textit{How can GenAI assistants facilitate code comprehension?}}
\begin{figure}[h!]
    \centering
    \includegraphics[width=\textwidth]{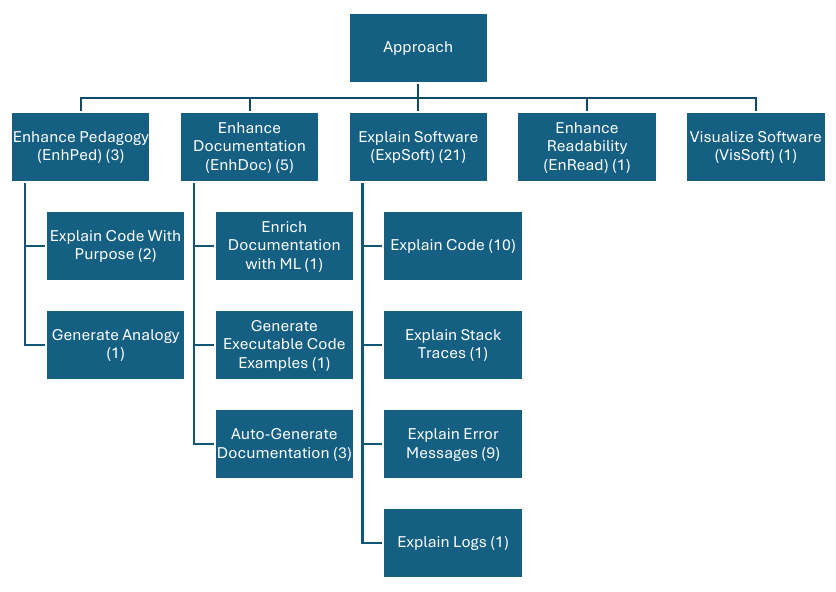}
    \caption{Approaches to enhancing code comprehension using GenAI. Number of papers categorized into each approach is shown in parentheses.}
    \label{fig:GenAI approach}
\end{figure}

GenAI assistants support code comprehension in five complementary ways:
explaining software, improving code readability, enhancing pedagogy through AI‐driven instructional interventions,
enhancing documentation, and visualizing software. We discuss these five approaches (Figure \ref{fig:GenAI approach}) in Section 4.1, their methods for measuring effectiveness in Section 4.2, and the effectiveness results in Section 4.3.

\subsubsection{Explain Software}
\paragraph{\textbf{Explain Code}}
Several tools have been developed to generate explanations to help programmers understand code.  CodeAid \cite{kazemitabaar2024codeaid}, implements an "Explain Code" feature through an interactive interface. Students can type or paste the code they want the tool to explain. After submission, CodeAid uses GenAI to provide students with real-time comprehension support, allowing them to hover over each line of code to receive detailed explanations. GILT \cite{nam2024using} can access the user's local code context and prompt the GenAI to generate explanations for the highlighted code without requiring the user to craft prompts . Ivie \cite{yan2024ivie} generates anchored explanations for GenAI-generated code by providing multi-level explanations. Based on previous empirical findings that developers are often unaware of code provenance, Tang et al. \cite{tang2024towards} developed a JetBrains plugin to enhance developers' understanding of GenAI-generated code through multi-level summarization. Developers can then edit these summaries based on their intent and use the edited summaries as prompts for the GenAI to refactor unsatisfactory code. 

Since GenAI predicts the next token in an input (a.k.a. the prompt), input quality significantly affects the output quality \cite{macneil2023experiences}. \textit{Prompt engineering} is the practice of crafting effective prompts to elicit solutions to the users' problems. Several studies have proposed novel prompt engineering approaches to enhance the code explanations generated by GenAI. One example is Ahmed et al.'s \cite{ahmed2024automatic} Automatic Semantic Augmentation of Prompts (ASAP), which integrates semantic information derived from code content into the prompt to improve code summarization. Similarly, Geng et al. \cite{geng2024large} leverages in-context learning to generate multi-intent comments based on developers' diverse perspectives in practice. Other studies have evaluated the quality of GenAI-generated code explanations for computing students \cite{macneil2023experiences, sarsa2022automatic, nguyen2024using, leinonen2023comparing}.

\paragraph{\textbf{Explain Error Message}}
Novices often struggle with programming error messages (PEMs) due to their poor readability, excessive jargon, and inadequate explanations \cite{santos2023always,cucuiat2024feedback}, which can lead to frustration \cite{santos2024not}. Researchers have begun exploring the use of GenAI to enhance existing PEMs or generate explanations for faulty code, with the goal of of improving developers' understanding of bugs and assisting them in debugging \cite{taylor2024dcc, widyasari2024demystifying, santos2023always, kang2024quantitative}. FuseFL \cite{santos2023always} generates explanations and fixes for code, given the erroneous code and the PEM. To enhance the explainability of spectrum-based fault localization (SBFL), Widyasari et al. \cite{widyasari2024demystifying} enhance GenAI responses to a code description coupled with SBFL outcomes, enabling the GenAI to offer developers concise reasoning about why specific lines of code are considered faulty. Similarly, DocHelp \cite{taylor2024dcc}, which is integrated into the Debugging C Compiler (DCC), refines compiler and runtime error messages by prompting a GenAI with the source code, error locations, and compiler error messages. AutoFL \cite{kang2024quantitative} automatically generates explanations using a GenAI to clarify error messages and describe the intent behind functions. Amburle et al. \cite{amburle2024ai} introduce an AI-based error explainer that leverages Google’s Gemini model to parse compiler and runtime error messages and generate concise, human-readable explanations for developers to streamline debugging. Other studies purely evaluated the quality of GenAI-enhanced error messages for students \cite{balse2023evaluating, cucuiat2024feedback, santos2024not, leinonen2023using}.

\paragraph{\textbf{Explain Stack Traces}}
Stack traces—sequences of method calls leading to an error—can be challenging for developers to interpret, particularly in unfamiliar codebases. To reduce the extensive human effort required for debugging and onboarding newcomers, Balfroid et al. \cite{balfroid2024towards} leverage generative AI to automate stack trace explanations. To create so-called ``code tours,'' they extract stack traces from failing tests and prompt the model to generate concise, step-by-step explanations for each relevant code segment, with the aim of enhancing developers' ability to quickly identify and resolve issues.

\paragraph{\textbf{Explain Logs}}
One study explored the use of GenAI to analyze software logs online to help comprehend complex software systems (e.g., distributed file systems, high-performance computing systems). Liu et al. \cite{liu2024interpretable} introduced LogPrompt to enhance the interpretability of log analysis, thereby supporting code comprehension in complex software systems.

\subsubsection{Enhance Pedagogy}
\paragraph{\textbf{Explain Code}}
 With the rise of GenAI, students need to devote increasing amounts of time to understanding and evaluating GenAI-generated code \cite{denny2024explaining}. Pedagogical interventions for code comprehension aim to equip students with the skills to understand code's purpose and functionality \cite{macneil2023experiences, leinonen2023comparing}. Denny et al. \cite{denny2024explaining} and Smith et al. \cite{smith2024prompting} proposed a novel approach in which students were asked to explain a given code snippet. Their explanations were then input into an LLM to determine whether the generated code matched the original. This method aimed to en students' code comprehension while addressing the subjectivity in evaluating students' written explanations. 
 
 \paragraph{\textbf{Generate Analogies}}
 Similarly, Bernstein et al. \cite{bernstein2024like} explored whether students could use GenAI to generate analogies to help them understand recursion functions.

\subsubsection{Enhance Documentation}
\paragraph{\textbf{Enrich Documentation with Machine Learning}}
One study enriched code documentation using Machine learning. Nam et al. \cite{nam2023improving} introduced a machine-learning-based knowledge extraction approach (SOREL) to automatically extract comparable APIs and explanatory sentences from Stack Overflow to help developers better understand unfamiliar APIs.

\paragraph{\textbf{Generate Executable Code Examples}}
One study utilized GenAI to generate executable code examples to supplement the code documentation. Khan et al. \cite{khan2023combining} proposed a novel approach that feeds the OpenAI Codex model with source code and natural language descriptions to generate executable code examples, with the aim of enhancing existing code documentation.

\paragraph{\textbf{Auto-Generate Documentation}}
Some studies leverage GenAI to generate code documentation automatically. To assist data scientists in crafting code documentation, Wang et al.  \cite{wang2022documentation} proposed Themisto, a deep-learning-based approach for generating code summaries automatically while still allowing users to refine them manually. In contrast, some research has focused on leveraging GenAI to enhance existing code documentation. Two studies generated and evaluated the quality of GenAI-generated code documentation using Codex and other advanced GenAI models \cite{dvivedi2024comparative, khan2022automatic}.

\subsubsection{Enhance Readability}
GenAI has been used to generate more understandable and readable function names, helping developers better understand the code \cite{nazari2024generating}. Nazari et al. \cite{nazari2024generating} introduce a GenAI-based technique for generating explanatory names for intermediate functions by providing input-output pairs. The generated function names are dual-validated by a program verifier and a secondary GenAI before being presented to developers.


\subsubsection{Visualize Software}
Software visualization uses visual representations and animations to assist developers in understanding how algorithms and code work \cite{bernstein2024like, heidrich2023visualizing, stasko1998software}. In the only work to leverage GenAI for software visualization, Heidrich et al. \cite{heidrich2023visualizing} employed Stable Diffusion \cite{rombach2022high} to generate comics illustrating source code, with the goal of making the software structure more comprehensible and easier to understand.

\subsection{RQ2: \textit{What methods have been used to study the use of GenAI in facilitating code comprehension?}}
We now review the methods used to study the use of GenAI assistance in code comprehension (RQ2). At the top level, we discuss dependent measures (the types of data collected), data collection methods (the methods used to collect data), data analysis methods (the methods used to analyze data), and participants (the individuals from whom the data is collected).

\subsubsection{Dependent Measures (DM)}
Based on Figure \ref{fig:countsofcategory}, we categorize dependent measures into the perception of GenAI-generated responses, the effectiveness of the GenAI application, reflections on pedagogy, human metrics, code readability metrics, code understanding metrics, and NLP metrics. We show subjective and objectiv dependent measures, respectively, in Table \ref{tab:subjective_data} and Table \ref{tab:objective_data}.

\paragraph{\textbf{Percept (Per)}}
Thirty one studies in our corpus collected subjective assessments of AI-generated explanations, documentation, and feedback. For example, learners and experts rated the overall clarity, accuracy, and effectiveness of GenAI‐generated code explanations \cite{macneil2023experiences, widyasari2024demystifying}. In a prompt‐engineering study, six experienced Java developers evaluated GenAI‐generated comments for naturalness, adequacy, and usefulness \cite{geng2024large}. Similarly, six professionals evaluated the interpretations of online logs using proposed prompt strategies in detail, specificity, relevance, logical soundness, and general helpfulness \cite{liu2024interpretable}. To compare human‐ and GPT‐generated explanations, Leinonen et al. \cite{leinonen2023comparing} evaluated differences in understandability, accuracy, and ideal versus actual length, and also asked an open‐ended question—“What is it about a code explanation that makes it useful for you?”—conducting a thematic analysis of 100 student responses.

Researchers have also assessed GenAI‐generated documentation. One study used metrics such as accuracy, completeness, relevance, understandability, and readability \cite{dvivedi2024comparative}; another applied a three‐dimensional rubric (readability, accuracy, informativeness) alongside self‐reported satisfaction scores \cite{wang2022documentation}. Participants in Nazari et al’s study \cite{nazari2024generating} judged the appropriateness of function names—followed by describing each subroutine’s functionality and articulating their overall understanding of the code—to assess the impact of GenAI‐generated names.

In work on code tour explanations, GenAI outputs were classified by transparency, scrutability, and efficiency \cite{balfroid2024towards}. CS instructors evaluated GenAI-enhanced error messages using binary scores for syntactic correctness, completeness, accuracy, and comprehensibility \cite{leinonen2023using, balse2023evaluating}. Similar rubric-based binary ratings were applied to feedback from GPT-4, focusing on error identification and the inclusion of model solutions \cite{nguyen2024using}. Manual ratings of AutoFL explanations considered accuracy, precision, conciseness, and usefulness \cite{kang2024quantitative}, while GenAI‐enhanced error messages were evaluated for conceptual accuracy, inaccuracy, relevance, and completeness \cite{taylor2024dcc}. Santos et al. \cite{santos2023always} rated sentence structure, clarity of explanations, and correction quality. To capture instructor perspectives, Cucuiat et al. \cite{cucuiat2024feedback} conducted semi‐structured interviews with eight educators to assess the quality of LLM‐generated feedback and explanations. Finally, Widyasari et al. \cite{widyasari2024demystifying} employed the BLEURT metric—a BERT‐based measure—to compare FuseFL‐generated explanations with human‐written ones.

Several experimental studies further analyzed GenAI outputs. Khan et al. \cite{khan2023combining} executed code examples produced by a GenAI model, recording execution success as a viability metric and checking each example’s relevance to the target method and consistency with provided documentation. Sarsa et al. \cite{sarsa2022automatic} examined 20 GenAI‐generated explanations, categorizing error types and their frequencies across different priming programs, evaluating whether every code segment was addressed, and calculating the proportion of correctly explained lines.

Two studies measured the developers' perception of comprehension tools on effectiveness during programming tasks. In one study, participants assessed distraction levels, workload, usability, and limitations of a comprehension tool \cite{yan2024ivie}. Similarly, in the CodeAid study, students rated the tool’s usefulness and perceived value, and provided detailed explanations for their evaluations \cite{kazemitabaar2024codeaid}.

Three studies elicited students’ reflections on how GenAI‐assisted pedagogy improved their code comprehension. For example, participants responded to questions designed to assess the effects of GenAI-based instructional strategies on learning outcomes and code comprehension \cite{smith2024prompting, denny2024explaining}. Bernstein et al. \cite{bernstein2024like} investigated students’ experiences with an analogy-generation activity.

\paragraph{\textbf{Process Metrics (PrM)}}
Several studies logged user interactions to gather process metrics of objective usage. MacNeil et al.\ \cite{macneil2023experiences} recorded the timestamps when students opened and closed an explanation panel for a given code snippet, enabling calculation of total viewing time and number of views. Likewise, CodeAid \cite{kazemitabaar2024codeaid} captured detailed interaction logs to identify which features students used most frequently and to characterize overall usage patterns. Taylor et al. \cite{taylor2024dcc} recorded each instance of erroneous source code, including the error’s location, the raw C compiler message, and ChatGPT’s response, and logged all student activities to compile comprehensive usage statistics. Nam et al. \cite{nam2023improving} assessed the participants' awareness and understanding of the differences between the comparable API methods. Finally, Wang et al. \cite{wang2022documentation} tracked several interaction metrics for their treatment group, such as clicks on the suggestion light bulb icon, direct use of generated documentation, manually authored documentation entries, and co-created documentation instances, and evaluated the quality of the final artifact by counting the number of Markdown cells and words added. 

\paragraph{\textbf{Performance Metrics (PeM)}}
Eight experimental studies measured human performance metrics to evaluate the effectiveness of GenAI-based approaches. For example, in the Ivie study \cite{yan2024ivie}, after completing programming tasks, participants answered 20 yes/no questions about an unfamiliar OpenCV API call, with both response times and accuracy recorded. Similarly, in the GILT study \cite{nam2024using}, participants completed an API‐related quiz to assess depth of understanding, and task performance was evaluated by both completion time and accuracy. In a laboratory experiment, Santos et al.\cite{santos2024not} recorded the time taken by participants to debug programs using GenAI-enhanced error-message explanations. Nam et al.\cite{nam2023improving} computed multiple objective variables—including task completion time, number of search queries issued, number of web pages visited, and solution correctness—to gauge the impact of AI-enhanced code documentation. Finally, Wang et al.\cite{wang2022documentation} also used task completion time as an indicator of performance in their documentation study.

Three studies employed eye-tracking to capture participants’ gaze behavior. In the Ivie study \cite{yan2024ivie}, attention was measured in terms of fixation duration on the generated code. Madi et al. \cite{al2022readable} developed a Visual Studio Code extension that tracked eye movements by computing metrics such as fixation count, total fixation time, first-fixation duration, and single-fixation duration to provide insights into cognitive processing.

\paragraph{\textbf{NLP Metrics (NM)}}
Five studies employed automated NLP metrics to assess the outputs or applications of GenAI models. To evaluate the quality of GenAI-generated explanations, Widyasari et al. \cite{widyasari2024demystifying} used the BLEURT metric, a BERT-based measure for natural language generation, to compare FuseFL-generated explanations with human-authored ones. In prompt-engineering research, Codex’s performance on code summarization was evaluated on two code-comment-generation datasets using BLEU, ROUGE-L, and METEOR metrics \cite{geng2024large}. Liu et al. \cite{liu2024interpretable} assessed LogPrompt on real-world log datasets across nine domains using the F1 score. Ahmed et al. \cite{ahmed2024automatic} evaluated the ASAP approach on the CodeSearchNet dataset using BLEU-CN, BLEU-DC, ROUGE-L, and METEOR metrics. Additionally, GenAI-generated code documentation has been analyzed with metrics such as documentation length and Flesch–Kincaid Grade Level to measure the volume and readability of the generated content \cite{khan2022automatic}.

\subsubsection{Data Collection Methods (DCM)}
\begin{figure}[h!]
    \centering
    \includegraphics[width=0.7\textwidth]{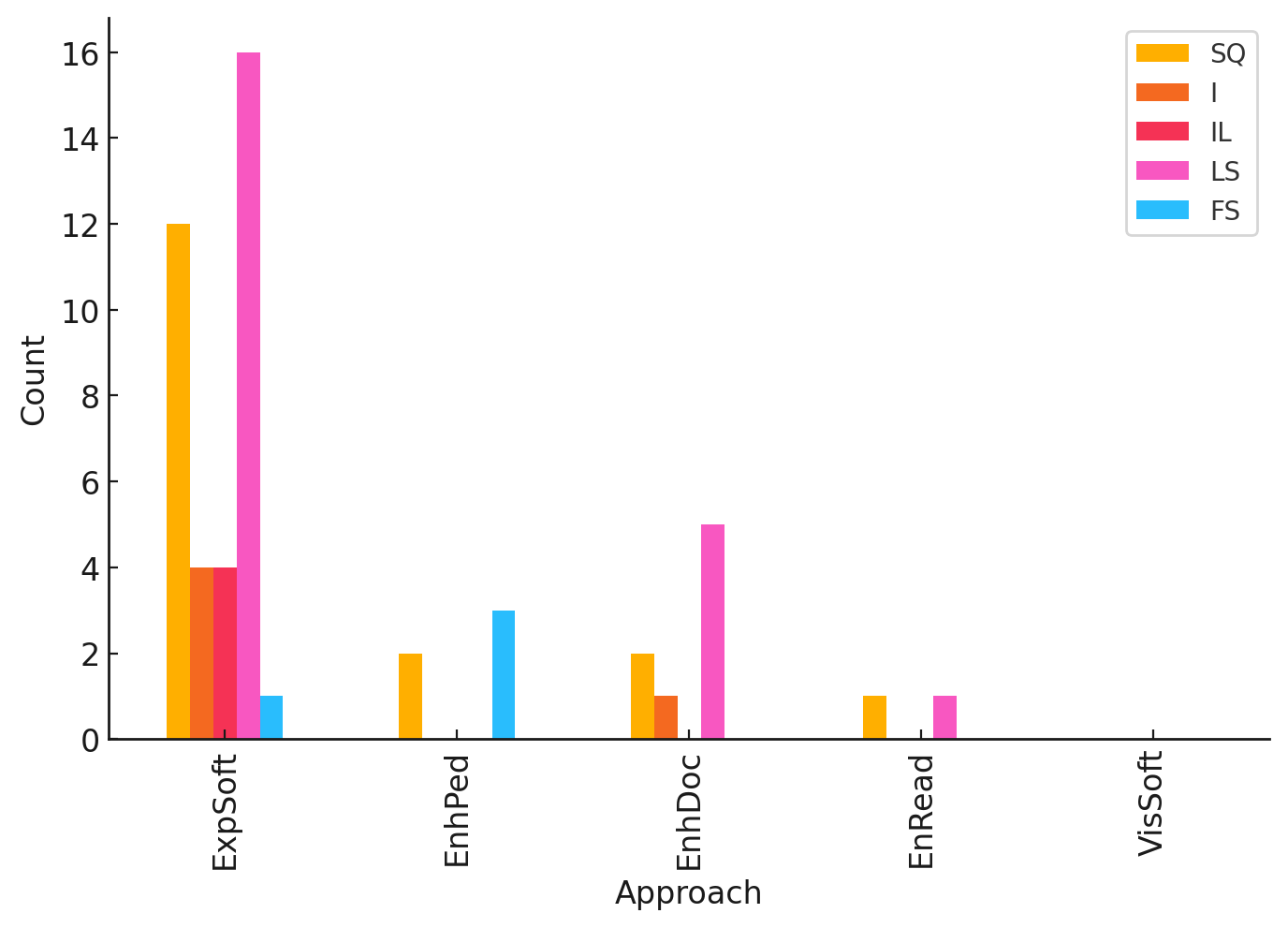}
    \caption{Data Collection Methods by Approach}
    \label{fig:DCM}
\end{figure}

Figure~\ref{fig:DCM} presents how many papers in each of the five top-level approaches employ each data collection method: Survey/Questionnaire (SQ), Field Study (FS), Lab Study (LS), Interview (I), and Interaction Logging (IL).

Lab studies are by far the most common method in Explain Software (ExpSoft=16) studies and are also used significantly in studies on Enhance Documentation (EnhDoc=5) and EnRich Documentation (EnRead=1), while they are entirely absent from studies on Enhance Pedagogy (EnhPed) and Visualize Software (VisSoft). Surveys follow a similar distribution: ExpSoft=12, EnhPed=2, EnhDoc=2, EnRead=1, and none in VisSoft. Field studies appear only in ExpSoft=1 and EnhPed=3. Interviews occur in ExpSoft=4 and EnhDoc=1, and interaction logging is confined to ExpSoft=4.

\subsubsection{Data Analysis Methods (DAM)}
\begin{figure}[h!]
    \centering
    \includegraphics[width=0.7\textwidth]{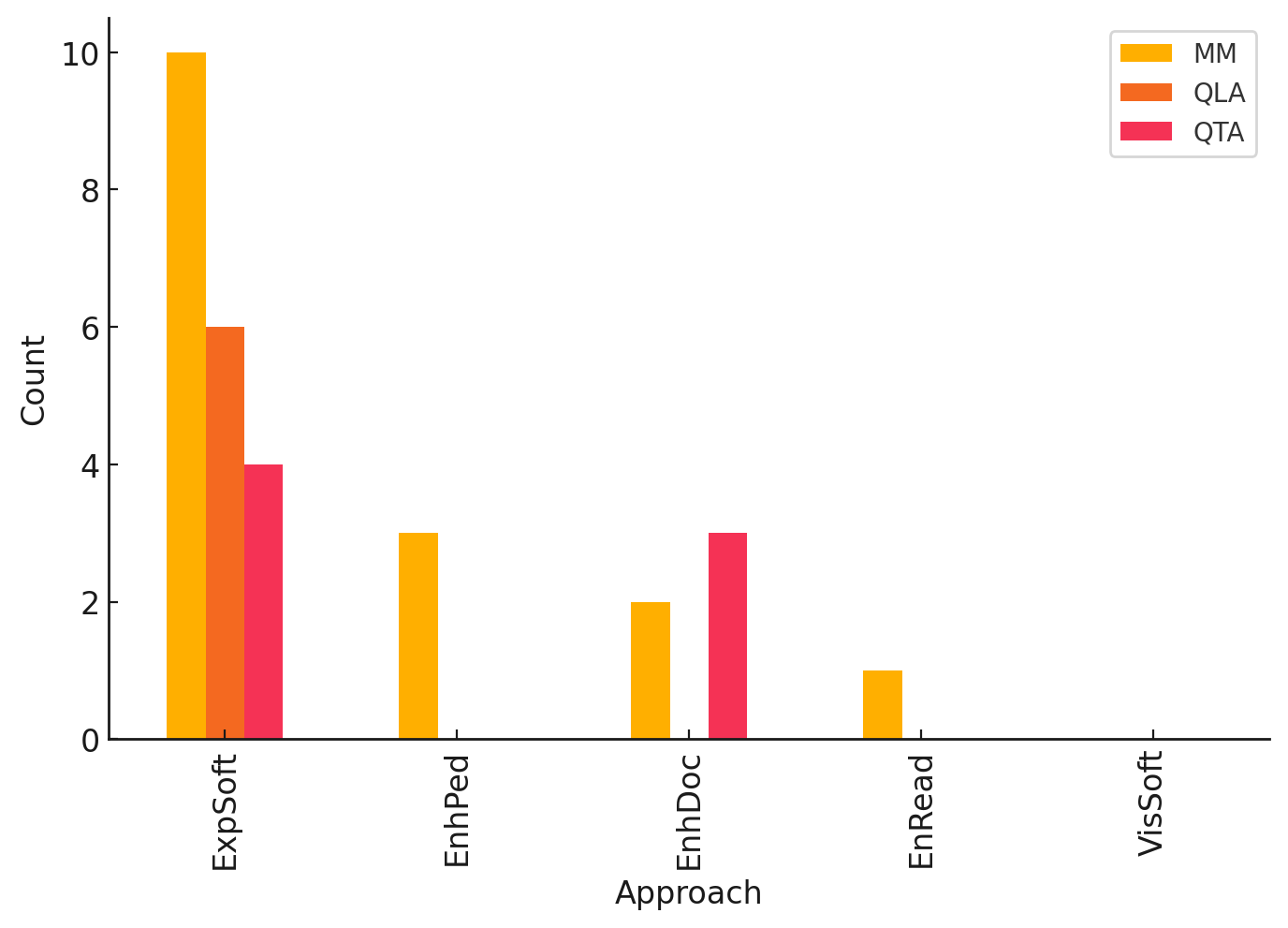}
    \caption{Data Analysis Methods by Approach}
    \label{fig:DAM}
\end{figure}

Figure~\ref{fig:DAM} shows the count of papers using Mixed Methods (MM), Qualitative Analysis (QLA), and Quantitative Analysis (QTA) within each approach.

Mixed methods dominate across four approaches—ExpSoft=10, EnhPed=3, EnhDoc=2, and EnRead=1—and are unused in VisSoft. Qualitative analysis is used exclusively in ExpSoft=6. Quantitative analysis appears in ExpSoft=4 and EnhDoc=3 only.

\subsubsection{Participants}
\begin{figure}[h!]
    \centering
    \includegraphics[width=0.7\textwidth]{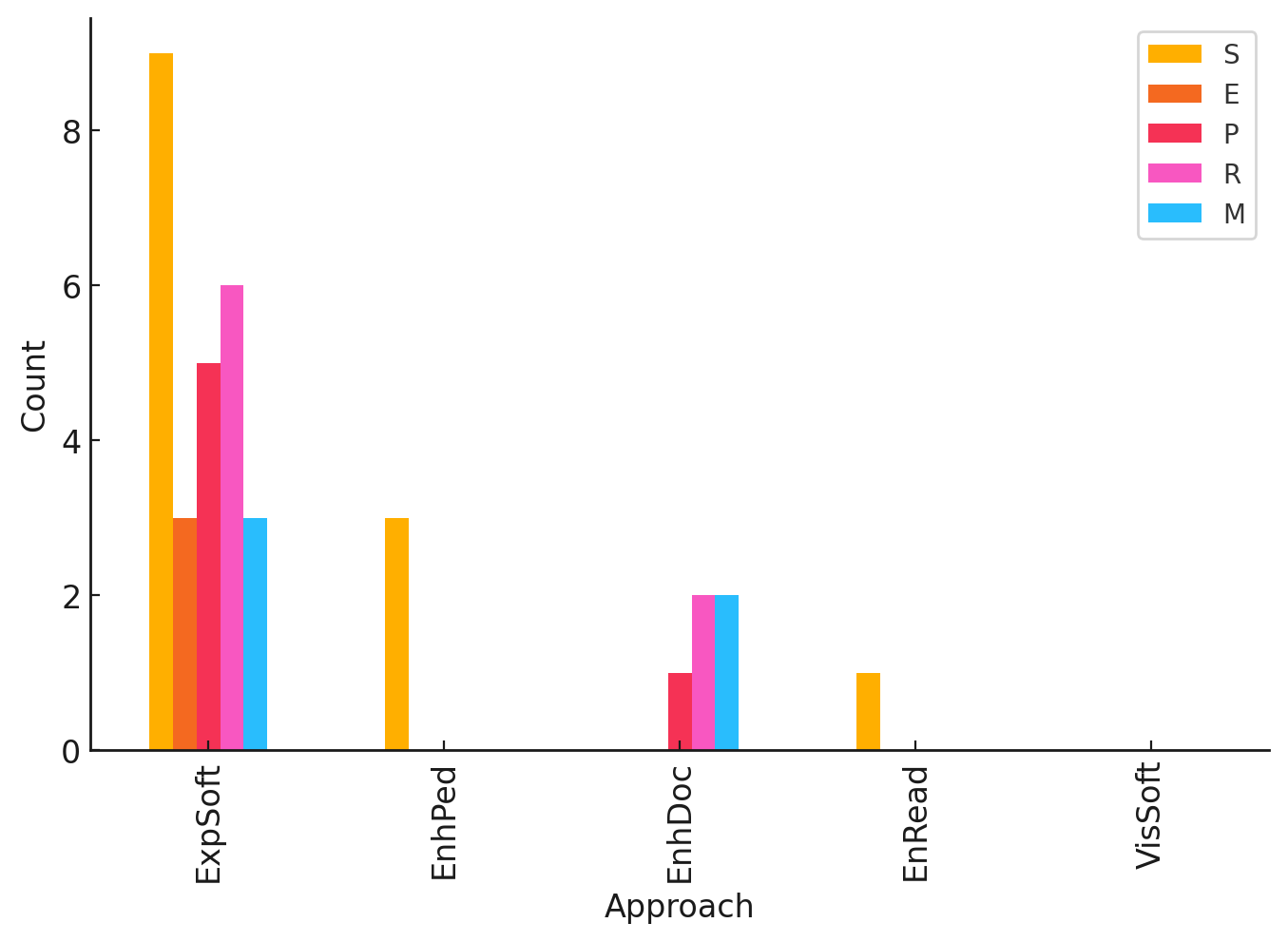}
    \caption{Participants by Approach}
    \label{fig:P}
\end{figure}

Figure~\ref{fig:P} breaks down participant types—students (S), educators (E), professionals (P), researchers (R), and GenAI models (M)—across each approach.

Explain Software engages all five types (S=9, E=3, P=5, R=6, M=3). Enhance Pedagogy involves only students (S=3). Enhance Documentation includes professionals (P=1), researchers (R=2), and models (M=2). EnRich Documentation features only students (S=1). Visualize Software reports no human or model participants.

\begin{table}[h!]
\footnotesize
\centering
\caption{Summary of Subjective Dependent Measures}
\label{tab:subjective_data}
\begin{tabular}{p{1.2cm} p{8.5cm} p{2cm}}
\hline
\textbf{DM} & \textbf{Explanation}                                            & \textbf{References}                                                \\ 
\hline
\multirow{24}{*}{Percept}   & Ease of use, Workload, Understanding Facilitating             & \cite{yan2024ivie,kazemitabaar2024codeaid}                        \\
   & Comparison                                                    & \cite{kazemitabaar2024codeaid}                                    \\
  & Usability, Limitations                                        & \cite{yan2024ivie}                                                \\
  & Experiences, Values                                           & \cite{kazemitabaar2024codeaid}                                    \\
  & Awareness, Understanding                                      & \cite{nam2023improving}                                           \\
 & Clarity, Correctness, Effectiveness                           & \cite{leinonen2023using,balse2023evaluating,geng2024large,liu2024interpretable,macneil2023experiences} \\
  & Accuracy, Completeness, Relevance, Understandability, Readability  & \cite{dvivedi2024comparative}                                     \\
   & Readability, Accuracy, Informativeness                        & \cite{wang2022documentation}                                      \\
   & Usability, Accuracy, Trustworthiness, Effectiveness           & \cite{wang2022documentation,nguyen2024using}                      \\
  & Appropriateness                                               & \cite{nazari2024generating}                                       \\
  & Understandability, Accuracy, Ideal Length, Actual Length      & \cite{leinonen2023comparing}                                      \\
  & Each subroutine’s functionality and overall code understanding & \cite{nazari2024generating}                                       \\
  & Usefulness                                                    & \cite{leinonen2023comparing}                                      \\
  & Transparency, Scrutability, Efficiency                        & \cite{balfroid2024towards}                                       \\
  & Syntactic correctness, Overall correctness, Conceptual accuracy & \cite{phung2023generating,kang2024quantitative,taylor2024dcc}    \\ 
   & Completeness of whether a response fully addresses the problem  & \cite{phung2023generating,taylor2024dcc}                          \\
   & Clarity, Comprehensibility, Explanations of errors, Fixes     & \cite{phung2023generating,leinonen2023using}                     \\ 
  & Conciseness, Usefulness                                       & \cite{kang2024quantitative}                                       \\
 & Error analysis                                                & \cite{balse2023evaluating}                                       \\
 & Relevance, Appropriateness                                    & \cite{taylor2024dcc}                                              \\
 & Sentence structure, Has explanation, Explanation correct, Has fix, Fix quality, Fix correct & \cite{santos2023always}    \\
 & Reflection                                                    & \cite{denny2024explaining,smith2024prompting}                    \\
   & Perception                                                    & \cite{denny2024explaining,smith2024prompting}                    \\
   & Experiences                                                   & \cite{bernstein2024like}                                         \\
\hline
\end{tabular}
\end{table}

\begin{table}[h!]
\footnotesize
\centering
\caption{Summary of Objective Dependent Measures}
\label{tab:objective_data}
\begin{tabular}{c p{7cm} c}
\hline
\textbf{DM} & \textbf{Explanations} & \textbf{Reference} \\
\hline
\multirow{7}{*}{Performance Metrics}   & Accuracy of comprehension questions/quizs & \cite{yan2024ivie,nam2024using} \\
     & Task completion time                         & \cite{nam2024using,santos2024not,nam2023improving,wang2022documentation} \\
     & Task completion accuracy                    & \cite{nam2024using,nam2023improving} \\
     & Number of search queries                    & \cite{nam2023improving} \\
     & Number of web pages visited                 & \cite{nam2023improving} \\
     & Fixation duration                           & \cite{yan2024ivie,al2022readable} \\
     & Fixation count, total fixation time         & \cite{al2022readable} \\ \hline
\multirow{5}{*}{Process Metrics}   & Viewing time of explanation                 & \cite{macneil2023experiences} \\
     & Number of viewing explanation               & \cite{macneil2023experiences} \\
     & Frequency of feature usage                  & \cite{kazemitabaar2024codeaid} \\
     & Student activities                          & \cite{taylor2024dcc} \\
     & Number of button clicks                     & \cite{wang2022documentation} \\
\hline
\multirow{8}{*}{NLP Metrics}   & BLEURT                                        & \cite{widyasari2024demystifying} \\
     & BLEU                                          & \cite{geng2024large} \\
     & ROUGE-L, METEOR                               & \cite{geng2024large,ahmed2024automatic} \\
     & F1 Score, precision, recall                   & \cite{liu2024interpretable,sahu2024codequeries,zeng2022extensive} \\
     & BLEU-CN                                       & \cite{ahmed2024automatic} \\
     & BLEU-DC                                       & \cite{ahmed2024automatic} \\
     & BLEU-4                                        & \cite{zeng2022extensive} \\
     & Exact Match                                   & \cite{sahu2024codequeries} \\
\hline
\multirow{5}{*}{Percept}  & Documentation length, Flesch–Kincaid grade level & \cite{khan2022automatic} \\
     & Passibility, relevance                           & \cite{khan2023combining} \\
     & Frequency of error, proportion of lines correctly explained & \cite{sarsa2022automatic} \\
     & Variety of analogy topics               & \cite{bernstein2024like} \\
     & Percentage of students completing tasks & \cite{smith2024prompting,denny2024explaining} \\
\hline
\end{tabular}
\end{table}

\subsection{RQ3: \textit{How effective are GenAI assistants in facilitating code comprehension?}}
We now present results that address the effectiveness of GenAI assistants using the methods discussed in RQ2. 
\subsubsection{Explain Software}
\paragraph{\textbf{Explain Code}}
Several studies have developed GenAI-powered tools to explain code, significantly enhancing users' code comprehension as confirmed by statistical tests. For example, in Yan et al. \cite{yan2024ivie}, programmers answered 90.2\% of comprehension questions correctly with Ivie—a tool that provides lightweight, anchored AI-generated explanations of just-generated code—versus 65.0\% with a baseline GPT-based in-editor chatbot, a 25.2\% improvement ($F = 23.6, p < 0.001$). They also responded more quickly using Ivie ($F = 9.82, p < 0.01$) and gave it a perfect mean self-reported comprehension rating (7 / 7) compared to the baseline. In Nam et al. \cite{nam2024using}, participants using GILT (Generation-based Information-support with LLM Technology) completed on average 0.47 more subtasks than when using a traditional search engine ($p < 0.01$), with professionals gaining 0.57 additional subtasks ($p < 0.01$), although students showed no significant improvement. These results indicate that while GenAI-based comprehension tools improve code understanding overall, the magnitude of benefit depends on the developer’s level of experience. The “Explain Code” feature of CodeAid \cite{kazemitabaar2024codeaid} generated code explanations assessed to be accurate 95\% of the time and perceived as beneficial, with a mean usefulness rating of 4.17 ($SD$ = 1.21). Conversely, Tang et al. \cite{tang2024towards} did not evaluate the effectiveness of their proposed tools in helping developers comprehend code.

Some studies have leveraged different prompt strategies to enhance the quality of GenAI-generated code explanations, aiming to improve code comprehension. Ahmed et al. \cite{ahmed2024automatic} used prompt engineering to enhance GenAI-generated code summaries across six programming languages, with BLEU scores (an automatic metric for evaluating the quality of machine-translated text) increasing from 1.84 (8.12\%) to 4.58 (16.2\%). Additionally, the results showed that the most effective prompt components are repository information, data flow graphs (DFG), and identifiers. Among these components, repository information significantly contributed to the effectiveness of using few-shot learning, which embeds a small number of input–output examples directly within the prompt to guide a GenAI model’s behavior ~\cite{brown2020language} ~\cite{ahmed2024automatic}. According to Geng et al. \cite{geng2024large}, a Codex-based model employing 10-shot learning (embedding 10 input-output examples within the prompt) with semantic demonstration selection and token-based reranking outperformed the state-of-the-art supervised approach DOME in generating multi-intent code comments. To further assess quality, Geng et al. \cite{geng2024large} conducted a human evaluation, which confirmed that comments with higher automatic metric scores also received higher participant ratings.

Some studies specifically evaluated the quality of GenAI-generated code explanations. For example, Leinonen et al.  \cite{leinonen2023comparing} found that GenAI-generated explanations were significantly more understandable than student-generated explanations, suggesting that GenAI-generated explanations could serve as scaffolding for students who are not proficient in creating their own explanations \cite{leinonen2023comparing}. 
Similarly, to assess whether GenAI can generate useful code explanations for students, MacNeil et al. \cite{macneil2023experiences} presented students with various types of GenAI-generated code explanations, such as line-by-line explanations, summaries, and concept listings, within an online e-book and asked them to rate these explanations. The results indicate that students found GenAI-generated explanations useful for understanding code and preferred summary explanations the most. Moreover, Sarsa et al. \cite{sarsa2022automatic} explored the quality of GenAI-generated line-by-line explanations and found that 90\% of code explanations covered all parts of the code, and 67.2\% of explanations were correct. Based on these findings, GenAI-generated explanations may provide a useful starting point for helping students understand or debug their code, despite often containing minor inaccuracies. 

Balse et al.~\cite{balse2023evaluating} evaluated GenAI’s ability to explain logic errors by asking teaching assistants to rank six explanations—five generated by teaching assistants and one generated by an LLM—for buggy code snippets. They found that the GenAI-generated explanation was frequently ranked among the top three most helpful. Their analysis showed that 93\% of GenAI-generated explanations contained at least one correct statement, 50\% included at least one incorrect statement, and 33\% omitted at least one logical error, indicating that LLM-generated explanations should be reviewed before being presented to students. Nguyen et al. \cite{nguyen2024using} experimented in which GPT-4 correctly assessed students’ submissions on conceptual understanding (92.04\%), syntax (89.38\%), and time complexity (90.27\%). Additionally, 92.03\% of its explanation suggestions were accurate, though 4.42\% included syntax errors. Computer science instructors and tutors judged these GenAI explanations to be highly useful, awarding 90.27\% for the clarity of code suggestions and 88.50\% for follow-up hints, which demonstrably improved students’ understanding of their code and the underlying programming concepts. However, GPT-4 occasionally flagged non-existent errors, mirroring findings from prior GenAI research \cite{nguyen2024using}.

\begin{tcolorbox}[colback=gray!10, colframe=black!70, 
                  width=\textwidth, arc=3mm, auto outer arc,
                  boxrule=0.5pt, drop shadow, fonttitle=\bfseries, 
                  title=Summary of GenAI's Effectiveness in Explaining Code]
GenAI-powered code comprehension tools consistently improve developers' understanding of code, though benefits vary by experience level, with professionals benefitting more than students. Enhancements in prompt engineering improved BLEU scores for GenAI-generated code summaries, while Codex models employing semantic demonstration selection surpassed traditional supervised methods in generating effective multi-intent code comments. Studies also indicate that GenAI-generated explanations are more understandable than those produced by students and highly valued by learners; however, accuracy remains imperfect, with frequent minor errors necessitating human review.
\end{tcolorbox}

\paragraph{\textbf{Explain Stack Traces}}
Balfroid et al. \cite{balfroid2024towards} evaluated the quality of GenAI-generated stack trace explanations (i.e., ``code tours'') in terms of transparency, scrutability, and efficiency. With rspect to transparency, GenAI did not explain some business terms related to the source code. For scrutability, although hallucination is a known issue with GenAI, no such instances were found in their evaluation. In terms of efficiency, the model often repeated readily available information, with 71\% of instances containing predefined keywords (e.g., param, return, etc.) \cite{balfroid2024towards}.

\begin{tcolorbox}[colback=gray!10, colframe=black!70, 
                  width=\textwidth, arc=3mm, auto outer arc,
                  boxrule=0.5pt, drop shadow, fonttitle=\bfseries, 
                  title=Summary of GenAI's Effectiveness in Explaining Stack Traces]
A study showed that GenAI-generated stack trace explanations lacked transparency due to missing business term clarifications, were free from hallucinations (scrutability), but showed limited efficiency by frequently repeating predefined keywords.
\end{tcolorbox}

\paragraph{\textbf{Explain Error Messages}}
GenAI models (e.g., Codex and GPT-4) can generate explanations for program error messages (PEMs) that are more comprehensible than the original messages, even though the issues they highlight vary depending on the error type and available code context \cite{leinonen2023using, santos2023always, kang2024quantitative, taylor2024dcc}. In particular, Codex generated comprehensible explanations in 67–100\% of cases and achieved accuracy rates of 11–83\% in identifying logic errors. However, although it suggested actionable fixes, only about half were correct; overall performance improved when using a lower temperature setting—i.e., reduced randomness—in the model’s responses \cite{leinonen2023using}.

Furthermore, Santos et al.'s study demonstrated that including code context significantly enhanced the quality of GenAI-generated explanations and fixes, with the perfect-fix rate increasing from 11\% without code context to 78\% with code context. This finding underscores the importance of contextual information in understanding code errors \cite{santos2023always}. Evaluations of FuseFL indicated that its generated explanations primarily focused on fixes, whereas human-generated explanations tended to emphasize diagnosis. Nevertheless, the overall clarity and informativeness of FuseFL-generated explanations were comparable to those produced by humans, with novice developers generally rating them as helpful \cite{widyasari2024demystifying}. 

Additionally, AutoFL produced correct explanations for the root causes of bugs in approximately 20\% of cases and provided at least one correct explanation for 56.7\% of bugs. However, developers noted issues with content overlap and inaccuracies, and they expressed a preference for more templated, concise feedback \cite{kang2024quantitative}. Finally, Doc--help generated explanations that performed notably better for compile errors, displaying up to 90\% conceptual accuracy and achieving tutor-level performance in 72\% of cases compared to runtime error explanations, which were less accurate, complete, and consistent, and sometimes included extra code blocks despite prompts instructing otherwise \cite{taylor2024dcc}. These findings indicate that while GenAI-based approaches can significantly enhance traditional error messages by providing concise, more actionable fixes and context-aware explanations, further refinement is needed to address technical and formatting issues \cite{taylor2024dcc}.

\begin{tcolorbox}[colback=gray!10, colframe=black!70, 
                  width=\textwidth, arc=3mm, auto outer arc,
                  boxrule=0.5pt, drop shadow, fonttitle=\bfseries, 
                  title=Summary of GenAI's Effectiveness in Explaining Error Messages]
GenAI models such as Codex and GPT-4 substantially enhance the comprehensibility and usefulness of program error messages, with their effectiveness varying by error type and code context. Accuracy and quality significantly improve with context inclusion (perfect-fix rate rising from 11\% to 78\%) and lower randomness settings; however, challenges persist, including occasional inaccuracies, redundant content, and format inconsistencies, highlighting the need for further refinement to reach human-level clarity and precision.
\end{tcolorbox}

\paragraph{\textbf{Explain Logs}}
LogPrompt \cite{liu2024interpretable} achieved the highest F1-score, which measures the harmonic mean of precision and recall, in six of the eight datasets for the log parsing task (extracting common and unique segments from raw logs), outperforming LogPPT \cite{le2023log} and LogStamp \cite{tao2022logstamp} by 32.8\% and 37.4\%, respectively. In the anomaly detection task (identifying anomalies in historical log sequences), LogPrompt improved the F1-score by 55.9\% over existing methods. In a human evaluation, expert reviewers blindly rated 200 LogPrompt outputs (100 for log parsing and 100 for anomaly detection) on usefulness and readability using a five-point scale, yielding average scores above four and High Interpretability Percentages (the proportion of samples receiving scores higher than four) exceeding 80\%.

\begin{tcolorbox}[colback=gray!10, colframe=black!70, 
                  width=\textwidth, arc=3mm, auto outer arc,
                  boxrule=0.5pt, drop shadow, fonttitle=\bfseries, 
                  title=Summary of GenAI's Effectiveness in Explaining Software Logs]
LogPrompt \cite{liu2024interpretable} substantially improved log parsing performance, outperforming prior approaches (LogPPT and LogStamp) by over 30\% in F1-score across multiple datasets, and boosted anomaly detection accuracy by 55.9\%, while also receiving strong human evaluations with over 80\% of outputs rated highly for usefulness and readability.
\end{tcolorbox}

\subsubsection{Enhance Pedagogy}
In the lab studies by Denny et al.\ \cite{denny2024explaining} and Smith et al.\ \cite{smith2024prompting}, 80\% of students agreed that this GenAI-enhanced pedagogy accurately evaluates code comprehension skills. Qualitative analysis revealed that the second most prevalent open-ended response was that the pedagogy improved students’ understanding of code. Similarly, only 63.1\% of students in Bernstein et al.'s \cite{bernstein2024like} study can successfully generate analogy-based explanations using GenAIs to aid code comprehension. However, some concerns remained that GenAI can impair students' understanding of code due to hallucinations in which LLM outputs contain factually incorrect, misleading, or entirely fabricated information \cite{wikipedia_hallucination_ai}.

\begin{tcolorbox}[colback=gray!10, colframe=black!70, 
                  width=\textwidth, arc=3mm, auto outer arc,
                  boxrule=0.5pt, drop shadow, fonttitle=\bfseries, 
                  title=Summary of GenAI's Effectiveness in Enhancing Pedagogy]
Most students found GenAI-enhanced pedagogy helpful in improving code comprehension and articulating code purpose, though inaccurate GenAI-generated code analogies may lead to misconceptions, further hindering code comprehension.
\end{tcolorbox}

\subsubsection{Enhance Readability}
The NomNom system \cite{nazari2024generating} achieved an overall accuracy of approximately 79\% in generating function names, compared to only 24\% with the baseline approach (GPT-3.5 and Code2Vec applied directly to raw synthesized code). Furthermore, in human studies, 76\% of developers noted that the names generated by NomNom were appropriate and helpful for understanding the code. In contrast, only 2\% of developers agreed that the names produced by the baseline approach were useful \cite{nazari2024generating}.

\begin{tcolorbox}[colback=gray!10, colframe=black!70, 
                  width=\textwidth, arc=3mm, auto outer arc,
                  boxrule=0.5pt, drop shadow, fonttitle=\bfseries, 
                  title=Summary of GenAI's Effectiveness in Generating Enhanced Function Names]
Nazari et al.'s study found that NomNom outperformed a baseline approach in generating function names and was deemed more helpful by developers for code comprehension.
\end{tcolorbox}

\subsubsection{Enhance Code Documentation}
GenAI models are now capable of generating code comments and explanations that match or even surpass original, human‐written documentation, making it easier for developers to understand and navigate unfamiliar codebases. In particular, closed‐source systems like GPT-3.5, GPT-4, and Bard consistently outperformed open‐source alternatives, especially when generating inline (function‐level) documentation that directly clarifies code behavior. Their file‐level summaries remained useful but tended to be less detailed than the granular, in‐context explanations these models provide \cite{dvivedi2024comparative}. Evaluations of Codex demonstrated that its performance varies across different programming languages. Moreover, employing one-shot learning improves Codex's BLEU scores—--surpassing state-of-the-art methods. Its documentation was found to be comparable to human-written documentation in terms of reliability and informativeness \cite{khan2022automatic}. Similarly, 72.5\% of GenAI-generated code examples compiled successfully, and 82.5\% of them effectively leveraged the target method and documentation, demonstrating the potential of GenAI-generated code examples to supplement code documentation and assist developers in understanding library functions \cite{khan2023combining}.  

In a human study of the Themisto system for data scientists, combining deep-learning-based and query-based documentation methods (i.e., a human-AI co-creation approach) significantly improved developers' productivity and their ability to understand unfamiliar code and APIs despite some of the generated explanations being vague and inaccurate \cite{wang2022documentation}. Likewise, SOREL (a machine-learning-based API for knowledge extraction) revealed superior performance in identifying the comparable APIs compared to the baselines and prior work \cite{nam2023improving}.

\begin{tcolorbox}[colback=gray!10, colframe=black!70, 
                  width=\textwidth, arc=3mm, auto outer arc,
                  boxrule=0.5pt, drop shadow, fonttitle=\bfseries, 
                  title=Summary of Effectiveness of GenAI-enhanced Eode Eocumentation]
\begin{itemize}
    \item Most large language models—particularly closed-source ones like GPT-3.5, GPT-4, and Bard—consistently generate documentation and code examples comparable or superior to human-written documentation, with performance varying by programming language.
    \item One-shot learning (e.g., for Codex) and human-AI co-creation (e.g., Themisto) can further enhance developer productivity and comprehension despite occasional inaccuracies.
\end{itemize} 
\end{tcolorbox}

%% file: sections/s5-discussion.tex
\subsection{Discussion}
We now discuss our findings relative to each RQ.
\subsubsection{RQ1: How can GenAI assistants facilitate code comprehension?}
Our review of 31 recent papers reveals that GenAI assistants support code comprehension through five main approaches: explaining software, enhancing pedagogy, generating documentation, enhancing/evaluating code readability, and visualizing software.

Explaining software involves providing code snippet explanations, clarifying error messages, interpreting stack traces, and analyzing software logs. Among these approaches, GenAI tools like CodeAid and Ivie can be effective in enhancing comprehension by providing on‐demand, contextualized explanations, yet occasional inaccuracies and repetitions indicate the need for mechanisms that surface provenance and confidence (e.g., model‐based confidence scores or provenance links). Pedagogical enhancements leverage GenAI both as an assessment oracle—automatically grading student explanations—and as an analogy generator. While most students report better understanding when using GenAI to generate analogies, hallucination risks can introduce misconceptions, suggesting that pedagogical intervention must combine GenAI feedback with human review to mitigate inaccurate outputs. Documentation generation via GenAI models such as Codex and GPT‐4 accelerates the generation of inline summaries and example code, boosting developer productivity. However, variability across programming languages and occasional ambiguity underscore the importance of integrating human‐in‐the‐loop refinement (e.g., human‐AI co‐creation workflows as in Themisto) to ensure both accuracy and clarity. 

Code readability improvements, such as NomNom’s function‐naming technique, demonstrate that GenAI can produce naming conventions and refactorings that align with human expectations. Software visualization, though not evaluated in terms of effectiveness, shows promise in using generative image models to translate code structures into visual metaphors (e.g., comics), which may particularly benefit novices by providing high-level conceptual models of code. Future work should evaluate the comparative effectiveness of visual versus textual explanations across novice and expert developers using GenAI.

Overall, these four approaches highlight GenAI’s versatility in scaffolding different facets of comprehension—ranging from low‐level syntax and error understanding to high‐level conceptual overviews—while also illuminating critical areas for refinement around explanation fidelity, interface integration, and hybrid human‐AI workflows.

\subsubsection{RQ2: What methods have been used to study the use of GenAI in facilitating code comprehension?}
Our results show a predominance of lab studies (LS) (appearing in 25 of 31 papers), typically involving controlled experiments where participants completed programming tasks or debugged code. Lab studies provide precise measurements (e.g., task completion time, accuracy), but their findings may not generalize to real-world development scenarios. By contrast, field studies (FS) are rare, suggesting an opportunity to investigate GenAI in helping developers comprehend code in real‐world development settings over longer durations.

In terms of dependent measures, subjective perceptions (Per) dominate, with 26 papers collecting impressions of clarity, usefulness, and trust. While user perceptions are vital for adopting the GenAI-generated explanations, overreliance on surveys may lead to subujective bias. Future work should combine subjective ratings with objective metrics—such as software usage logs and performance outcomes—to triangulate effectiveness. Performance metrics (PeM) and process metrics (PrM) were used in eight and six studies, respectively. Although these metrics are often specific to particular tools/studies, future studies should apply those metrics to comprehensively evaluate the effectiveness of GenAI in helping code comprehension.

Regarding data analysis, a mixed methods (MM) approach was the most common, reflecting a growing number of studies that combine quantitative (e.g., task completion time, BLEURT scores) and qualitative methods (e.g., thematic coding). Purely quantitative (QTA) or qualitative (QLA) analyses appear less frequently, suggesting that future studies should apply both methods.

Finally, participant demographics reveal a skew toward students (S), particularly in pedagogy and documentation research, with fewer studies involving professionals (P) or educators (E). As GenAI tools advance, inclusion of a broader range of developers over longer periods of time will be crucial to better understand GenAI's real‐world benefits and challenges in code comprehension.

\subsubsection{RQ3: How effective are GenAI assistants in facilitating code comprehension?}
Approaches to explaining software using GenAI have proven effective in various ways, albeit with certain limitations. GenAI tools such as GILT and Ivie consistently demonstrated improved performance across multiple dimensions of code comprehension, including comprehension accuracy, task completion speed, and subjective usefulness. Notably, significant performance improvements were particularly evident among professional developers, suggesting that experienced users may derive more immediate and measurable benefits from GenAI tools compared to novices or students. Consequently, future educational interventions should consider customized GenAI tools designed specifically to enhance benefits for less experienced learners. Enhanced prompt engineering techniques (e.g., few-shot learning \cite{wang2022documentation}) significantly improved the quality of GenAI-generated explanations, emphasizing the critical role of carefully crafted input prompts. Particularly effective strategies included incorporating context-specific details (e.g., repository information, data flow graphs, identifiers) and utilizing few-shot learning approaches \cite{ahmed2024automatic, geng2024large}. Future research should integrate established prompt design guidelines into computing education and practical development \cite{white2023prompt}. Furthermore, GenAI-generated explanations were consistently rated as beneficial and understandable, frequently exceeding the quality of explanations provided by less-experienced human users, such as students. However, the occurrence of minor inaccuracies, omissions, and occasional errors necessitates human oversight, indicating that GenAI should be integrated into hybrid workflows involving human reviewers rather than used independently. Finally, despite substantially improving the clarity and comprehensibility of error messages, GenAI-generated explanations still face persistent issues, including redundant content, inaccuracies, and formatting inconsistencies. Addressing these limitations remains crucial for the effective deployment of GenAI tools.

With respect to pedagogy, while GenAI-enhanced pedagogical strategies were generally well-perceived by students, there remains a critical risk of misunderstandings due to GenAI hallucinations. This finding underscores the necessity for educators to proactively identify and address potential misconceptions when integrating GenAI into educational contexts.

Studies on code documentation indicate that practitioners should leverage closed-source GenAI models due to their superior performance in generating high-quality documentation. The utilization of GenAI-generated documentation and executable code examples appears especially beneficial for developers dealing with unfamiliar APIs or legacy codebases. Furthermore, adopting a human-AI co-creation approach for generating code documentation appears promising and should be explored more extensively in future work.

Overall, frequent minor inaccuracies, occasional major errors, hallucinations, and redundant or repetitive information remain significant limitations that necessitate human oversight. Future research should focus on developing methods to mitigate these issues, such as enhanced prompting strategies, hybrid human-AI workflows, continuous human evaluation, and interactive approaches for model refinement to help developers/students comprehend code better.

\subsection{Implications}
We now consider the implications of our SLR's findings for computing students, instructors, researchers, and tool builders.
\subsubsection{For Computing Students}
The emergence of GenAI-based pedagogy, in which GenAI generates explanations to aid students in understanding code, marks a shift in code comprehension within computing education. Students are now actively engaged in understanding GenAI outputs, suggesting that future computing education research should focus on the comprehension skills needed to evaluate, debug, and explain these outputs. As several studies note, while GenAI-generated explanations can enhance understanding, they may also include inaccuracies or uncertainties (hallucinations) that could mislead students. Consequently, students must develop critical skills to assess the correctness of GenAI outputs, especially as these tools become more prevalent. Moreover, students should be cautious about over-relying on GenAI-generated code, as excessive dependence might diminish their ability to comprehend code independently.

\subsubsection{For Computing Instructors}
Instructors should consider deploying GenAI assistants in the classroom in cases where detailed, line-by-line explanations and summary views of code are needed to scaffold students' comprehension process. Moreover, tools like CodeAid can be used to help break down complex code into understandable segments, enabling students to form a clear mental model of code functionalities. Such scaffolding reduces students' cognitive load of crafting the prompts requesting explanations from GenAI models. Instructors also can use GenAI assistance to generate formative feedback to allow students to identify gaps in their comprehension and learn from AI-generated feedback. 

Instructors can help students develop a deeper understanding of code by prompting them to critically evaluate AI-generated explanations, verify code correctness, and engage in reflective activities. In addition, instructors should set appropriate guardrails by requiring students to answer comprehension questions about the AI-generated code before copying and pasting it.

Furthermore, instructors can use RAG \cite{lewis2020retrieval} to integrate course-specific knowledge (e.g., lecture notes and curated code examples) into prompts that enhance AI-generated code explanations, as this holds promise in improving students' code comprehension. In addition, instructors should learn how to fine-tune pre-trained models to automatically assess students' self-explanations of code purpose, as fine-tuned models have demonstrated high performance in grading students' self-explanations \cite{chapagain2025automated}.

\subsubsection{For Computing Education Researchers}
Computing education researchers should design controlled experiments that compare GenAI-assisted interventions with traditional learning methods, incorporating longitudinal studies to determine whether short-term gains in comprehension translate into sustained improvement. Researchers should focus on metrics such as comprehension accuracy, time on task, and cognitive load, employing both qualitative methods (e.g., thematic analysis of student responses) and quantitative measures (e.g., performance scores and eye-tracking data) to assess the impact of GenAI on code comprehension comprehensively. Furthermore, insights from empirical studies can be leveraged to build theoretical models that explain how and why GenAI tools affect code comprehension. These models can inform further research and provide a basis for designing more effective GenAI-enhanced code comprehension environments.

\subsubsection{For GenAI Tool Developers}
Tools such as GILT and Ivie demonstrate that integrating local code context into prompts can significantly improve the accuracy and usefulness of GenAI-generated explanations, thereby reducing the cognitive load on users when crafting prompts. Therefore, tool builders should ensure that GenAI-based comprehension tools capture as much contextual information as possible, whether through IDE plugins or automatic code extraction, to generate accurate responses.

Most current GenAI-based comprehension tools capture contextual information only from a single code chunk, and none utilize GenAI to incorporate historical information from repositories (e.g., commit messages, lines changed over time). Although the work by Horvath et al. \cite{horvath2024meta} tracks historical information to help developers comprehend an unfamiliar code base, it does not leverage the power of GenAI.
Furthermore, while proactive assistance (e.g., automatic error analysis) can reduce cognitive load and speed up comprehension, it may also hinder developers’ problem-solving and comprehension skills. Similar to educational tools for code comprehension, a key challenge for tool developers is to find the right balance between unsolicited, context-driven help and user-initiated interactions.
Lastly, although GenAI-generated explanations are helpful, issues such as hallucinations, inaccuracies, and redundant context generation persist. Therefore, tool builders should develop robust validation mechanisms (e.g., human-in-the-loop checks, confidence estimates, or cross-verification with secondary GenAI models) to enhance trust and usability.

Another approach that tool builders should consider is to integrate advanced prompt engineering strategies into code comprehension tools to improve the consistency and quality of code and repository explanations. Additionally, incorporating user feedback into explanation refinement may help reduce errors and better align GenAI-generated explanations with developers’ expectations. Future tools should also allow developers to customize the level of detail, tone, and format of explanations to suit their expertise or the specific context of their coding tasks. For example, novice developers might benefit from detailed, step-by-step breakdowns, while experts may prefer concise code summaries. Seamless IDE integration can reduce context-switching and cognitive load by bridging the gap between GenAI-generated explanations and traditional development workflows.

\begin{table}[]
\small
\centering
\caption{Implications of SLR Results for Students, Instructors, Researchers, and Tool Developers}
\begin{tabular}{p{2cm} p{10cm}}
\hline
\textbf{Audience}   & \textbf{Implications} \\ 
\hline
Students            & 
\begin{itemize}[leftmargin=*]
  \item Need to develop skills to evaluate, debug, and explain AI-generated code.
  \item Be aware of potential inaccuracies (hallucinations) in AI explanations and avoid overreliance on AI outputs.
\end{itemize} \\ 
\hline
Instructors         & 
\begin{itemize}[leftmargin=*]
  \item Use AI assistants (e.g., CodeAid) to provide detailed, line-by-line code explanations and scaffold student learning.
  \item Leverage AI to generate formative feedback and prompt students to critically evaluate AI-generated explanations.
  \item Require students to answer comprehension questions before using AI-generated code.
  \item Incorporate course-specific knowledge via RAG for improved clarity.
\end{itemize} \\ 
\hline
Researchers         & 
\begin{itemize}[leftmargin=*]
  \item Run controlled, longitudinal studies that compare GenAI‐assisted code comprehension with traditional methods.
  \item Measure comprehension accuracy, time on task, and cognitive load using both quantitative (performance scores, eye‐tracking) and qualitative (student feedback) methods.
  \item Build theoretical models from empirical results to guide the design of more effective AI‐enhanced code‐comprehension tools.
\end{itemize} \\ 
\hline
Tool developers     & 
\begin{itemize}[leftmargin=*]
  \item Ensure tools capture rich local code context (via IDE plugins or automatic code extraction) to enhance explanation accuracy.
  \item Consider integrating repository historical information for deeper code comprehension.
  \item Strike a balance between proactive assistance and user-initiated interaction, and develop robust validation (human-in-the-loop, confidence estimates, etc.) to mitigate inaccuracies.
  \item Allow developers to customize the detail, tone, and format of explanations and ensure seamless IDE integration to reduce cognitive load.
\end{itemize} \\ 
\hline                                                    
\end{tabular}
\label{tab:implication}
\end{table}

%% file: sections/s6-threat.tex
\subsection{Internal threats}
There are four common threats to validity in systematic literature reviews (SLRs): paper selection bias, data extraction bias, use of incorrect search terms in digital libraries, and incomplete coverage of relevant journals and conferences \cite{feng2024guiding}. To mitigate paper selection bias, we adopted methodologies from previous studies \cite{feng2024guiding, stol2016grounded, heinonen2023synthesizing}. Specifically, we first identified search terms by reviewing prominent studies on code comprehension and then validated these terms through a pilot search across three widely recognized digital libraries. To address the threat of incomplete coverage, we employed backward snowballing, author snowballing, and a completeness check. Additionally, since data extraction for grounded theory is a manual process that may introduce bias, we conducted the open coding protocol between the first and second authors until thematic saturation was reached. 

\subsection{External threats}
A key external threat to this systematic literature review is that our search was confined to studies published from 2022 to 2024; as a result, any relevant papers released after our cut-off date will not be captured.

%% file: sections/s7-conclusion.tex
In this paper, we conducted a systematic literature review of state-of-the-art approaches and tools that leverage GenAI assistants to facilitate code comprehension. Our analysis of 31 papers published from 2022 to 2024 demonstrates promising innovations, including prompt engineering, GenAI-based pedagogy, and advanced comprehension tools. The literature included in our SLR indicates that GenAI assistants have the potential to mitigate the challenges of code comprehension faced by computer science students in programming classes and by practitioners during the maintenance phase. In summary, GenAI assistants are shifting coding activities toward a more comprehension-centric approach, although the methodologies and practical implementations of this transformation are still evolving. Future work should focus on developing standardized evaluation metrics for these methods and tools, enhancing the clarity of GenAI-generated explanations, and bridging the gap between GenAI-generated responses and the human cognitive process of code comprehension. Addressing these issues will be crucial for advancing both computing education research and practice to enhance code comprehension.